\ifx\mnmacrosloaded\undefined \input mn\fi
\input psfig

\def\lc{_{\rm lc}}
\def\c{_{\rm c}}
\def\js{J_{\rm s}} \def\jc{J_{\rm c}} \def\jt{J_{\rm lc}}
\def\Eh{\E_{\rm h}}
\def\Flc{F^{\rm lc}}
\def\Flcsol{F^{\rm lc}_\odot}
\def\Flcmf{F^{\rm lc}_{\rm MF}}
\def\Fdrain{F^{\rm drain}}
\def\Fdrainsol{F^{\rm drain}_\odot}
\def\Fdrainmf{F^{\rm drain}_{\rm MF}}
\def\Flw{F^{\rm lw}}
\def\Flwsol{F^{\rm lw}_\odot}
\def\Flwmf{F^{\rm lw}_{\rm MF}}
\def\Ftri{F^{\rm triax}}
\def\Fsol{F_\odot}
\def\Fmf{F_{\rm MF}}
\def\DR{\big\langle \Delta R \big\rangle}
\def\DRR{\big\langle (\Delta R)^2 \big\rangle}
\def\oa#1{#1_{\rm O.A.}}
\def\oa#1{\bar #1}
\def\iso#1{\bar #1}
\def\jl{J_l}
\def\d{{\rm d}}\def\p{\partial}
\def\E{{\cal E}}
\def\pr{\mathop{\smash{\rm p}\vphantom{\sin}}}
\def\DR{\big\langle \Delta R \big\rangle}
\def\DRR{\big\langle (\Delta R)^2 \big\rangle}
\def\oa#1{#1_{\rm O.A.}}
\def\oa#1{\bar #1}
\def\iso#1{\bar #1}
\def\jl{J_l}
\def\b#1{{\bmath#1}}

\def\halff{{\textstyle{1\over2}}}

\def\lta{\la}\def\gta{\ga}

\def\kms{{\rm\,km\,s^{-1}}}
\def\kpc{{\rm\,kpc}}
\def\Mpc{{\rm\,Mpc}}

\def\pc{{\rm\,pc}}

\def\yr{{\rm\,yr}}

\def\eqname#1{%
   \global\advance\Eqnno by1
   \xdef#1{(\theeq)}
   \global\advance\Eqnno by-1
}
\everydisplay{\puteqnum}
\def\puteqnum#1$${#1\eqno\stepeq$$}
\def\refeq#1{%
 \advance\Eqnno by -#1
 \advance\Eqnno by 1
 \hbox{(\theeq)}%
 \advance\Eqnno by #1
 \advance\Eqnno by-1
}

\def\ifundefined#1{%
   \expandafter\ifx\csname#1\endcsname\relax
}
\def\ref#1{\ifundefined{#1} $\bullet$#1$\bullet$%
  \immediate\write16{Undefined reference #1}%
            \write-1{Undefined reference #1}
\else\hbox{\csname#1\endcsname}\fi
}

\newif\ifAMStwofonts

\ifCUPmtplainloaded \else
  \NewTextAlphabet{textbfit} {cmbxti10} {}
  \NewTextAlphabet{textbfss} {cmssbx10} {}
  \NewMathAlphabet{mathbfit} {cmbxti10} {} 
  \NewMathAlphabet{mathbfss} {cmssbx10} {} 
  \ifAMStwofonts
    \NewSymbolFont{upmath} {eurm10}
    \NewSymbolFont{AMSa} {msam10}
    \NewMathSymbol{\upi}     {0}{upmath}{19}
    \NewMathSymbol{\umu}     {0}{upmath}{16}
    \NewMathSymbol{\upartial}{0}{upmath}{40}
    \NewMathSymbol{\leqslant}{3}{AMSa}{36}
    \NewMathSymbol{\geqslant}{3}{AMSa}{3E}

     \let\le=\leqslant
     
  \else
    \def\umu{\mu}
    \def\upi{\pi}
    \def\upartial{\partial}
  \fi
\fi

\pageoffset{-2.5pc}{0pc}

\loadboldmathnames




\begintopmatter  

\title{Rates of tidal disruption of stars by massive central black holes}
\author{John Magorrian$^{1,2}$ and Scott Tremaine$^{3}$}
\affiliation{$^1$ CITA, University of Toronto, 60 St George Street,
  Toronto, Ontario, Canada M5S 3H8}
\vskip0.1truecm
\affiliation{$^2$ Institute of Astronomy, Madingley Road, Cambridge
CB3 0HA}
\vskip0.1truecm
\affiliation{$^3$ Princeton University Observatory, Peyton Hall, 
Princeton, NJ 08544, USA}

\shortauthor{S. J. Magorrian and S. Tremaine}
\shorttitle{Tidal disruption rates}


\abstract{%
There is strong evidence for some kind of massive dark object in the centres
of many galaxy bulges.  The detection of flares from tidally disrupted stars
could confirm that these objects are black holes (BHs).  Here we present
calculations of the stellar disruption rates in detailed dynamical models of
real galaxies, taking into account the refilling of the loss cone of stars on
disruptable orbits by two-body relaxation and tidal forces in non-spherical
galaxies.  The highest disruption rates (one star per $10^4\yr$) occur in
faint ($L\lta10^{10}L_\odot$) galaxies, which have steep central density
cusps.  More luminous galaxies are less dense and have much longer relaxation
times and more massive BHs.  Dwarf stars in such galaxies are swallowed whole
by the BH and hence do not emit flares; giant stars could produce flares as
often as every $10^5\yr$, although the rate depends sensitively on the shape
of the stellar distribution function.  We discuss the possibility of detecting
disruption flares in current supernova searches.  
The total mass of stars consumed over the lifetime of the galaxy is of order
$10^6M_\odot$, independent of galaxy luminosity; thus disrupted stars may
contribute significantly to the present BH mass in galaxies fainter than $\sim
10^9L_\odot$.}


\maketitle  

\section {Introduction}

Both gas- and stellar-dynamical measurements indicate that many, if not most,
galaxies harbour some kind of massive central dark object (MDO) (e.g.,
Kormendy \& Richstone~1995; Magorrian et al.\ 1998; van der Marel 1998; Ford
et al.\ 1998; Ho~1998).  Indirect arguments suggest that these MDOs are most
probably black holes (BHs) since dark clusters of the required mass and size
are difficult to construct and maintain (Maoz~1998), and since dead quasar
engines should lurk in the centres of many nearby galaxies (So{\l}tan~1982;
Chokshi \& Turner~1992).  An inevitable source of fuel for these dead quasars
is the debris from tidal disruption of stars on almost radial orbits. Much of
the debris from a disrupted star gets ejected, but a portion remains bound to
the BH and emits a ``flare'' lasting a few months to a year
(Rees~1988, 1998; Lee 1999).  The spectrum is expected to be
mainly thermal, peaking at extreme UV or soft X-ray wavelengths.  Plausible
models (Ulmer~1998) predict $V$-band luminosities of $\sim10^9L_\odot$, and
even higher luminosities in the $U$ band (but see also Ulmer, Paczy\'nski \&
Goodman~1998).  Such flares ought to be easily detectable, but as yet there is
only marginal observational evidence that they occur (Komossa \& Bade~1999).

Recently Magorrian et al.\ (1998, hereafter Paper~I) used HST photometry and
ground-based kinematics to constrain MDO masses for a sample of 32 nearby
galaxies, assuming that each is well described by a two-integral axisymmetric
model.  The purpose of the present paper is to calculate the tidal disruption
rates for these two-integral models, assuming that their MDOs are BHs.  The
paper is organized as follows.  The following section gives a brief
description of our modelling assumptions and describes the ``loss cone'' of
stars on disruptable orbits.  In Section~3 we calculate the steady-state
disruption rate due to the diffusion into the loss cone of stars on
centrophobic loop orbits.  Similar calculations of this rate have recently
been presented by Syer \& Ulmer (1999).  In addition, in a flattened or
triaxial galaxy, a significant portion of phase space is occupied by
centrophilic orbits.  We investigate the effect these have on the disruption
rate in Section~4.  We then use these results to estimate the effects of
consumed stars on BH masses (Section~5) and to estimate the expected rate of
detection of flares in surveys (Section~6).  Before summing up, we discuss
possible shortcomings of our models in Section~7.

\section {General background: the loss cone}

Consider a galaxy with a central BH of mass $M_\bullet$.  Let us
suppose for simplicity that the galaxy is otherwise composed entirely
of stars of mass~$m_\star$ and radius~$r_\star$.  (The generalization
to a range of $m_\star$ and~$r_\star$ is dealt with in Appendix~A.)
Then, neglecting relativistic effects, a star that comes within a
distance
$$
r_{\rm t}=\left(\eta^2{M_\bullet\over m_\star}\right)^{1/3}r_\star
\eqname\tidalrad
$$
of the central BH will be tidally disrupted, where $\eta\simeq2.21$ for a
homogeneous, incompressible body and $0.844$ for an $n=3$ polytrope (Sridhar
\& Tremaine 1992, Diener et al.\ 1995).
We define the ``loss cone'' to consist of all orbits with pericentres
less than~$r_{\rm t}$. Of course, if $r_{\rm t}\lta r_{\rm
S}\equiv2GM_\bullet/c^2$ the star is swallowed whole by the BH,
without producing a flare.  Thus for $M_\bullet\gta10^8M_\odot$
solar-type stars are swallowed whole and only giants are disrupted
outside the BH's horizon.  However, it often proves convenient to
ignore the fact that the BH has a horizon so that results obtained for
some particular values of $r_\star$ and $m_\star$ can be scaled for
other values.  Thus in what follows, our ``consumption'' rates are the
rates at which stars come within a radius $r_{\rm t}$ of the BH, even
if this lies inside the BH's horizon.  In contrast we shall use the
term ``flaring'' rate to refer to the rate of disruption of stars
outside the BH's horizon.

Let us first examine the loss cone in a spherical galaxy.  The distribution
function (DF) $f(\b x,\b v)$ is defined such that $f(\b x,\b v)\,\d^3\b
x\d^3\b v$ is the probability of finding a star within a volume $\d^3\b
x\d^3\b v$ of $(\b x,\b v)$.  By Jeans' theorem, the DF of a spherical galaxy
depends on $(\b x,\b v)$ only through the stars' orbital binding energy per
unit mass~$\E=\psi(\b x)-{1\over 2}v^2$ and angular momentum per unit
mass~$J=|\b x \times \b v|$, where the relative gravitational potential
$\psi(r)$ is related to the galaxy's potential $\Phi(r)$ through
$\psi(r)\equiv-\Phi(r)$.  In terms of $(\E,J)$ the loss cone is given by
$$
J^2\le J\lc^2(\E)\equiv 2r_{\rm t}^2\left[\psi(r_{\rm t})-\E\right] \simeq
2GM_\bullet r_{\rm t},\eqname\LCboundary
$$
and we have assumed $\E\ll
GM_\bullet/r_{\rm t}$ (i.e. most stars are consumed from nearly radial
orbits).  The number of stars in any small volume of phase space $\d\E\d J^2$
around $(\E,J^2)$ is
$$\eqalign{
N(\E,J^2)\,\d\E\d J^2 & = \int \d^3\b x\,\d^3\b v\,f(\E,J^2)\cr
 & = 4\pi^2f(\E,J^2)P(\E,J^2)\,\d\E\d J^2,\cr
}\eqname\NEexpr
$$
where
$P(\E,J^2)$ is the radial period of an orbit with energy $\E$ and
angular momentum $J$.  For almost radial loss-cone orbits we may
approximate $P(\E,J^2)$ by $P(\E)\equiv P(\E,0)$.  So, for a spherical
galaxy with isotropic DF $f(\E)$, the number of stars $N\lc(\E)$ in the (full)
loss cone per unit energy is given by
$$
N\lc(\E)\,\d\E=4\pi^2 f(\E)P(\E)J^2\lc(\E)\,\d\E,
\eqname\NE
$$
while the total number of stars $N(\E)$ per unit energy interval
is given by
$$
N(\E)\,\d\E=4\pi^2 f(\E)\,\d\E\,\int_0^{J\c^2(\E)}P(\E,J^2)\,\d J^2,
\eqname\NlcE
$$
where $J\c(\E)$ is the angular momentum of a circular orbit of
energy~$\E$.

In an axisymmetric galaxy with an integrable potential the DF is a function
$f(\E,J_z,J_3)$, where $J_z$ is the component of angular momentum along the
symmetry ($z$-) axis and $J_3$ is a third integral.  Numerical experiments
show that we may reasonably make the approximation $J_3\simeq J$ (e.g., Binney
\& Tremaine 1987).  Then the number of stars in any small volume of
phase space $\d\E\d J_z\d J$ is just
$$N(\E,J_z,J)\,\d\E\d J_z\d J = 4\pi^2f(\E,J_z,J)P(\E,J_z,J)\,\d\E\d
J_z\d J.
\eqname\nnnnn
$$
If $P$ and $f$ are approximately independent of $J_z$ we can integrate over
$J_z$ from $-J$ to~$J$ and recover equation~\ref{NEexpr}.  Therefore, the
expressions~\ref{NE} and \ref{NlcE} given above also apply to axisymmetric
galaxies, at least for the interesting low-$J$ orbits.  In particular,
two-integral axisymmetric models correspond to spherical isotropic models if
we replace $f(\E)$ by $f(\E,J_z=0)$.  In what follows, we write $f(\E)$ for
the axisymmetric DF $f(\E,J_z=0)$, $f(\E,J^2)$ for $f(\E,J_z=0,J^2)$ etc.

For each of the two-integral models of Paper~I it is straightforward
to use the axisymmetric luminosity distribution $j(R,z)$ and
best-fitting BH mass~$M_\bullet$ and stellar mass-to-light
ratio~$\Upsilon$ to obtain $f(\E)$, $N(\E)$ and~$N\lc(\E)$.  The
method is as follows.  The best-fit stellar mass density to the
observations is $\rho(R,z)=\Upsilon j(R,z)$.  Assuming all stars are
of mass~$m_\star$, the corresponding number density $\nu(R,z)=\Upsilon
j(R,z)/m_\star$.  The $J_z=0$ slice of the DF then follows on
inverting 
$$\nu(0,z)=4\pi\int_0^{\psi(0,z)}
f(\E,J_z=0)\sqrt{2[\psi(0,z)-\E]}\,\d\E.
\eqname\eddington
$$
We solve~\refeq1 for $f(\E,J_z=0)$ using the numerical method
described in Appendix~C.

Fig.~1 shows the resulting $f(\E)$, $N(\E)$ and $N\lc(\E)$ for the
edge-on model of M32, assuming that it is composed entirely of
solar-type stars.  Both $f(\E)$ and $N\lc(\E)$ peak at
$\E\sim\Eh\equiv\psi(r_{\rm h})$, where $r_{\rm h}$ is the radius of
the sphere of influence of the BH, defined in terms of the galaxy's
intrinsic velocity dispersion~$\sigma(r)$ through 
$$\sigma^2(r_{\rm h})={GM_\bullet\over r_{\rm h}}\equiv\sigma_{\rm h}^2.
\eqname\infl
$$
The results depend on the assumed inclination angle of the galaxy, since
flatter galaxies have fewer stars on low-angular momentum orbits.  For this
edge-on model, the full loss cone contains about $15M_\odot$ of stars,
approximately half of which are bound to the BH.  Full loss cones in larger,
core galaxies contain up to about $2000M_\odot$.

\beginfigure{1}
\psfig{file=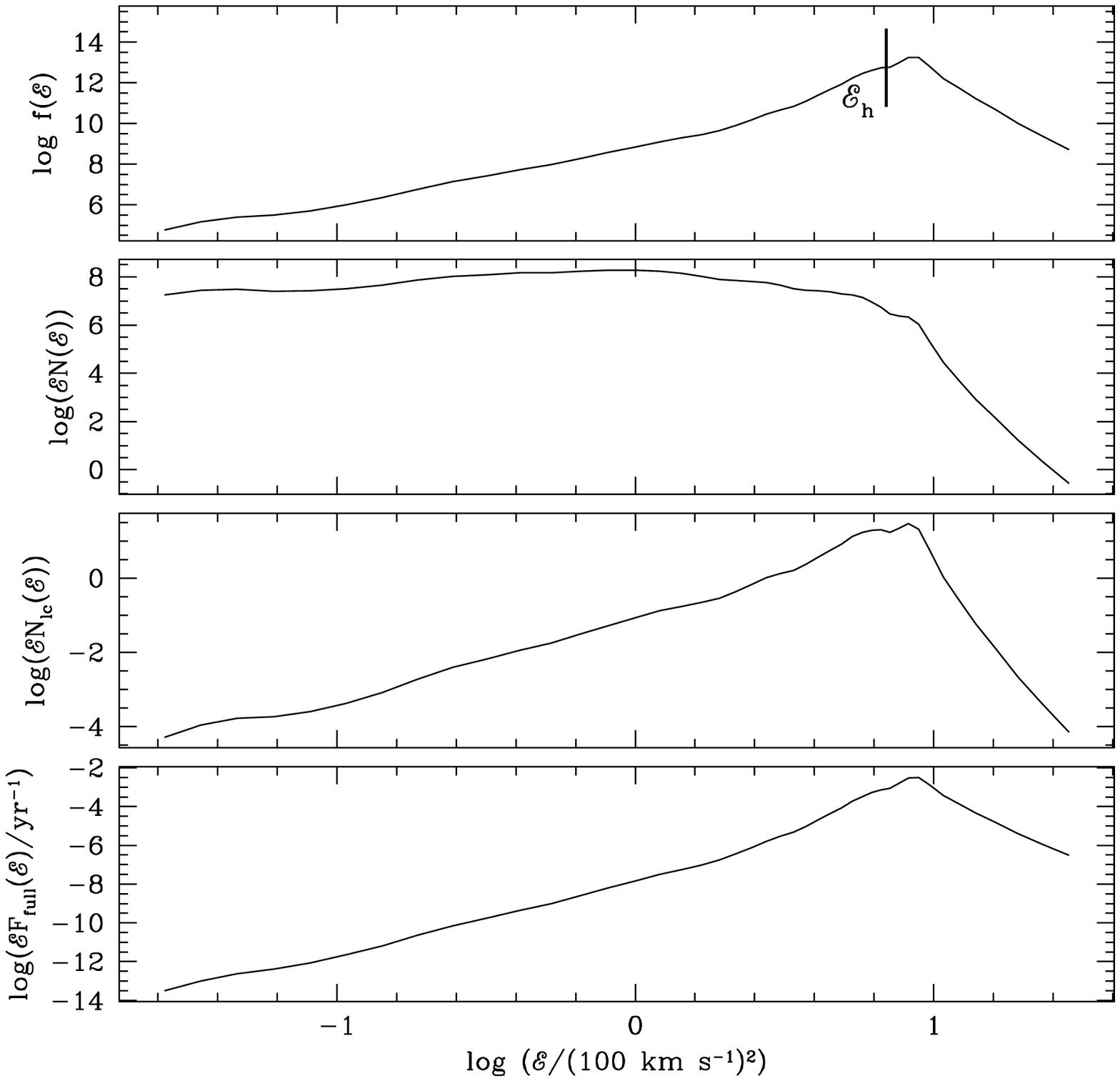,width=\hsize}
\caption{{\bf Figure 1.} Full loss cone in the edge-on model of M32.  The
top panel shows the isotropic DF $f(\E)$ obtained assuming the galaxy
is composed of solar-type stars.  The others show $N(\E)$ and
$N\lc(\E)$ given by equations \ref{NE} and~\ref{NlcE}, and the
consumption rate were the loss cone full, $F_{\rm
full}=N\lc(\E)/P(\E)$.  The heavy vertical line marks the
energy~$\E_{\rm h}=\psi(r_{\rm h})$, where $r_{\rm h}$ is the radius
of the sphere of influence of the BH (eq. \ref{infl}).}
\endfigure

The shapes of $f(\E)$, $N\lc(\E)$ and $N(\E)$ for $\E\gta\Eh$
can be understood as follows.  For $r\ll r_{\rm h}$,
$\psi=GM_{\bullet}/r$, $J\c^2(\E)=G^2M_\bullet^2/2\E$, and $P(\E)=2\pi
GM_\bullet/(2\E)^{3/2}$.  If the number density of stars is
$\nu(r)=\nu_0(r_{\rm h}/r)^{\alpha}$ then
$$f(\E)=(2\pi\sigma_{\rm h}^2)^{-{3\over2}}\nu_0
	{\Gamma(\alpha+1)\over\Gamma(\alpha-{1\over2})}
	\left(\E\over\sigma_{\rm h}^2\right)^{\alpha-{3\over2}},\quad
 {1\over2}<\alpha<3.
\eqname\DFcentral
$$
Equations \ref{NE} and~\ref{NlcE} then give for $\E\gta\Eh$
$$\eqalign{
N(\E) & \simeq {10^4\over(100\kms)^2}
             \left(M_\bullet\over10^6M_\odot\right)^3
             \left(\nu_0\over 10^5\pc^{-3}\right)\times\cr
&\qquad
             \left(100\kms\over\sigma_{\rm h}\right)^8
             \left(\E\over\sigma_{\rm h}^2\right)^{\alpha-4}\cr
N\lc(\E) & \simeq {0.5\over(100\kms)^2}
             \left(M_\bullet\over 10^6M_\odot\right)^{7/3}
             \left(\nu_0\over 10^5\pc^{-3}\right)\times\cr
&\qquad
             \left(100\kms\over\sigma_{\rm h}\right)^6
             \left(\E\over\sigma_{\rm h}^2\right)^{\alpha-3}
             \left(r_\star\over r_\odot\right)
             \left(m_\odot\over m_\star\right)^{1/3},\cr
}\eqname\NEnumeric
$$
both steeply falling functions of~$\E$. 
Thus $N(\E)$ and $N\lc(\E)$ decline at large $\E$, and $f(\E)$
declines if and only if $\alpha<3/2$.  For energies $\E\lta\Eh$ (radii
$r\gta r_{\rm h}$), the potential becomes a much shallower function of
radius, and so both $f(\E)$ and $N\lc(\E)$ fall as $\E$ is reduced
from $\E\sim\Eh$, whereas $N(\E)$ continues to rise due to its
dependence on $J\c^2$.  Thus $N\lc(\E)$ always peaks near $\Eh$, and
$f(\E)$ peaks near $\Eh$ in core galaxies. (M32 is the only
``power-law'' (Lauer et al.\ 1995) galaxy in our sample for which
$f(\E)$ declines at large $\E$. The reason is that the classification
of M32 as a power-law is based on its appearance at distances typical
for our galaxy sample, whereas in fact it is much closer. The density
profile of M32 is quite shallow within the innermost arcsecond or so,
a feature which would not be resolved if it were not so close.)

Since $N\lc(\E)$ peaks at $\E\sim\Eh\sim\sigma_{\rm
h}^2$, the characteristic timescale for emptying a full loss cone is
$$P(\Eh)\sim10^4\left(M_\bullet\over 10^6M_\odot\right)
\left(100\kms\over\sigma_{\rm h}\right)^3\yr.
$$
The flux of stars per unit energy into the BH's maw when the loss cone is full
is just $F^{\rm full}(\E)=N\lc(\E)/P(\E)$.  For $\E\gta\Eh$,
$$
\eqalign{
F^{\rm full}(\E) & \simeq {4\times10^{-5}\yr^{-1}\over(100\kms)^2}
             \left(M_\bullet\over 10^6M_\odot\right)^{4/3}
             \left(\nu_0\over 10^5\pc^{-3}\right)\times \cr
&\qquad
             \left(100\kms\over\sigma_{\rm h}\right)^3
             \left(\E\over\sigma_{\rm h}^2\right)^{\alpha-{3\over2}}
             \left(r_\star\over r_\odot\right)
             \left(m_\odot\over m_\star\right)^{1/3}.
\eqname\fulleq
}$$  Integrating
over $\E$, this reduces to the expression given by Rees (1988) when
$\alpha=0$.  For $\alpha>{1\over2}$ the total rate $\int F^{\rm
full}(\E)d\E$ diverges, being dominated by the consumption of the most
tightly bound stars.

\section {Repopulating the loss cone by two-body relaxation}

Two-body relaxation causes stars to diffuse gradually into a depleted
loss cone.  Stars diffuse in both $\E$ and $J^2$ with the same
characteristic timescale $t_{\rm relax}$, but diffusion in $J^2$ is
the dominant contributor to the consumption rate: for all but the most
tightly bound stars (of which there are very few -- see
eq.~\ref{NEnumeric}), the loss cone boundary is almost independent of $\E$
(eq.~\ref{LCboundary}).  Not all
stars that enter the loss cone are consumed: if the
characteristic change in $J^2$ per orbit $\Delta J^2\sim
J\c^2(\E)P(\E)/t_{\rm relax}(\E)\gg J\lc^2(\E)$, stars with $r\gg
r_{\rm t}$ can wander in and out of the loss cone many times per orbit
with impunity (Lightman \& Shapiro~1977).  

In this section we
calculate the steady-state rate of diffusion of stars into the loss
cone.  Our calculation is a simple generalization of Cohn \&
Kulsrud (1978; hereafter CK) to non-Keplerian potentials.
Let us define $f(\E,J^2;r)\d\E \d J^2$ to be the probability of finding a star
within $\d\E\d J^2$ of $(\E,J^2)$ at a given radius~$r$.  Neglecting
the effects of large-angle scatterings, the evolution of $f(\E,J;r)$
near the loss cone can be approximated by the Fokker--Planck equation
(e.g., CK)
$${\p f\over\p t}+v_r{\p f\over\p r}={\p\over\p R}
  \left[-\DR f + {1\over2}{\p\over\p R}\Big(\DRR f\Big)\right],
   \eqname\FPone
$$
where $R\equiv J^2/J^2\c(\E)$, $v_r=[2(\psi(r)-\E)]^{1/2}$ is the radial
velocity, and the diffusion coefficients $\DR$ and $\DRR$, both functions of
$(\E,R;r)$, measure the rate at which two-body encounters cause stars to
diffuse in~$R$.  We derive expressions for the diffusion coefficients in
Appendix~B, and find that $\DR={1\over2}\p\DRR/\p R$, simplifying the
right-hand side of~\ref{FPone}.  We seek a steady-state solution, so ${\p f/
\p t}=0$.  Expanding the diffusion coefficients about $R=0$, the
Fokker--Planck equation for $R\ll1$ becomes
$$
{\p f\over\p r}={\mu\over v_r}{\p\over\p R}\left(R{\p f\over\p R}\right),
\eqname\FPtwo
$$
where $\mu(\E,r)$ is the limiting value of $\DRR/2R$ as
$R\rightarrow0$, which is given
by $\mu=2r^2\langle\Delta v_t^2\rangle/J^2\c$, where $\langle\Delta
v_t^2\rangle$ is the diffusion coefficient for tangential velocity
(Appendix B).  Integrating over some small volume of phase space
$\d^3\b x\,\d^3\b v=8\pi^2J\c^2(\E)\,\d r\,{\d R}\,{\d\E}/v_r$ yields
the $(-R)$-directed flux of stars with energies between $\E$ and
$\E+\d\E$:
$$
\Flc(\E)\,\d\E = 8\pi^2J\c^2(\E)\,\d\E\, \int_{r_-}^{r_+} {\d r\over v_r}
\mu R{\p f\over\p R}.\eqname\eqflux
$$
Here $r_+$ and $r_-$ are the apo- and peri-centre radii of an orbit
with energy~$\E$ and angular momentum~$R$.

Our solution $f(\E,R;r)$ of~\ref{FPtwo} must satisfy two conditions.
First, since stars on loss-cone orbits are consumed at pericentre
$r=r_-$, we must have $f(\E,R;r_-)=0$ for $R<R\lc(\E)$. (Strictly
speaking, we should impose this condition for all $r<r_{\rm t}$.)
Second, since we seek a steady-state solution, $f(\E,R;r)$ should not
change from one orbital period to another.  To accommodate the latter
condition, we follow CK in changing variable from
$r$ to the time-like 
$$\tau\equiv
 \int_{r_-}^r{\mu\,\d r\over v_r}\Big/
   2\int_{r_-}^{r_+} {\mu\,\d r\over v_r}
 \equiv \int_{r_-}^r{\mu\,\d r\over v_r}\Big/P(\E)\oa{\mu}(\E),$$
so that $\tau=0,1,2,\ldots$ correspond to successive pericentre
passages of an imaginary star of energy $\E$ and angular momentum~$R$.
The variable $\oa\mu(\E)$ is just the orbit-averaged diffusion
coefficient.  The Fokker--Planck equation then becomes
$${\p f\over\p\tau}=P\oa\mu
   {\p\over\p R}\left(R{\p f\over\p R}\right),
\eqname\FP
$$
and our boundary conditions on $f(\E,R;\tau)$ are
$$\eqalign{
  f(\E,R;0) & = f(\E,R;1) \cr
  f(\E,R;0) & = f(\E,R;1) = 0\quad\hbox{if $R<R\lc(\E).$}\cr
}\eqname\FPbc
$$

CK point out that the orbit-averaged solution to
\ref{FP} subject to the boundary conditions~\ref{FPbc} is well approximated by
$$
f(\E,R)
\simeq A(\E)\ln\left(R\over R_0(\E)\right),\qquad R>R_0,
\eqname\twobodyDF$$
where $A(\E)$ is a constant and $R_0$ depends on
$q(\E)\equiv{P(\E)\oa\mu(\E)/R\lc(\E)}$ through
$$R_0(\E) = R\lc(\E)\times\cases{
      \exp(-q) & for $q(\E)>1$\cr
      \exp(-0.186q-0.824\sqrt{q}) & for $q(\E)<1$.\cr}
\eqname\rnought
$$
In the ``pinhole'' limit (Lightman \& Shapiro 1977) $q\gg1$, stars can wander
in and out of the loss cone many times without ill effect, and this is
reflected in the fact that the DF goes to zero only at $R_0\ll R\lc$. On the
other hand, in the ``diffusion'' limit $q\ll1$ stars cannot wander very far
into the loss cone without being consumed, so $R_0$ is very nearly equal to
$R\lc$.

Since most stars have $R_0(\E)\ll1$, the DF given by equation~\ref{twobodyDF}
is very close to the ``isotropized'' DF
$$\iso f(\E)\equiv A(\E)\int_{R_0(\E)}^1 \ln\left(R\over
R_0(\E)\right)\,\d R 
\simeq A(\E)\ln R_0^{-1}(\E).
$$
So, if a galaxy is observed to be consistent with an isotropic DF
$\iso f(\E)$, then its underlying DF, for $R>R_0(\E)$, is really
$$f(\E,R)={\iso f(\E)\over \ln R_0^{-1}(\E)}
\ln\left(R\over R_0(\E)\right).
\eqname\isoDF
$$
Substituting this into equation~\ref{eqflux} gives
the rate of consumption of stars by the BH,
$$
\Flc(\E)\,\d\E = 
4\pi^2 P(\E)J^2\c(\E)\oa\mu(\E){\iso f(\E)\,\d\E\over\ln R_0^{-1}(\E)}.
\eqname\twobodynetrate
$$
Using equation~\ref{rnought} we can rewrite this result as (see also
eq. (12) of Lightman \& Shapiro 1977) 
$$\Flc(\E)\,\d\E \sim \cases{
F^{\rm max}(\E)d\E/\ln\left(GM/4\E r_{\rm t}\right)
                                               & $q\ll-\ln R\lc$\cr
q^{-1}F^{\rm max}(\E)\d\E                & $q\gg-\ln R\lc$,\cr
}
\eqname\LSrate$$
where $F^{\rm max}(\E)\equiv4\pi^2 P(\E)J^2\c(\E)\oa\mu(\E)\iso
f(\E)\simeq N(\E)\oa\mu(\E)$ is an estimate of the maximum possible
flux through a constant-$\E$ surface in phase space (Lightman \&
Shapiro~1977).  Using the definition of~$q$ we can confirm that
$\Flc(\E)\simeq F^{\rm full}(\E)$ for $q\gg-\ln R\lc$, where $F^{\rm
full}$ is the consumption rate when the loss cone is full
(eq. \fulleq). Note that the transition between pinhole and diffusion
r\'egimes is at $q\simeq-\ln R\lc$ rather than $q\simeq 1$ as simpler
arguments would suggest (e.g. Lightman \& Shapiro 1977).

\subsection {Results}

Having $\iso f(\E)$ from equation~\ref{eddington}, it is straightforward to
calculate $\bar\mu(\E)$ and use equation~\ref{twobodynetrate} to calculate the
consumption rate for each of the galaxy models in Paper~I.  Fig.~2 shows the
results for our edge-on models of M32 and M87, assuming that the stars all
have solar mass and radius.  Note that in both cases $\Flc(\E)$ peaks at
$\E\simeq\Eh$, a result which we find holds generally to within about 20\%: at
$\E\simeq\Eh$, $\bar f(\E)$ turns over (Section~2), $\bar\mu(\E)$ levels off,
and the rates of change of both $J\c(\E)$ and $P(\E)$ peak.  Thus, while a
cursory glance at equation~\ref{LSrate} might suggest that $\Flc(\E)$ should
peak where $q\simeq-\ln R\lc$, it is the $F_{\rm max}(\E)$ factor that is the
dominant influence on the shape of $\Flc(\E)$.  Indeed, stars in large
galaxies like M87 just barely make it into the $q\gg1$ pinhole r\'egime and
never have $q\gg-\ln R\lc$.

\beginfigure*{2}
\centerline{\psfig{file=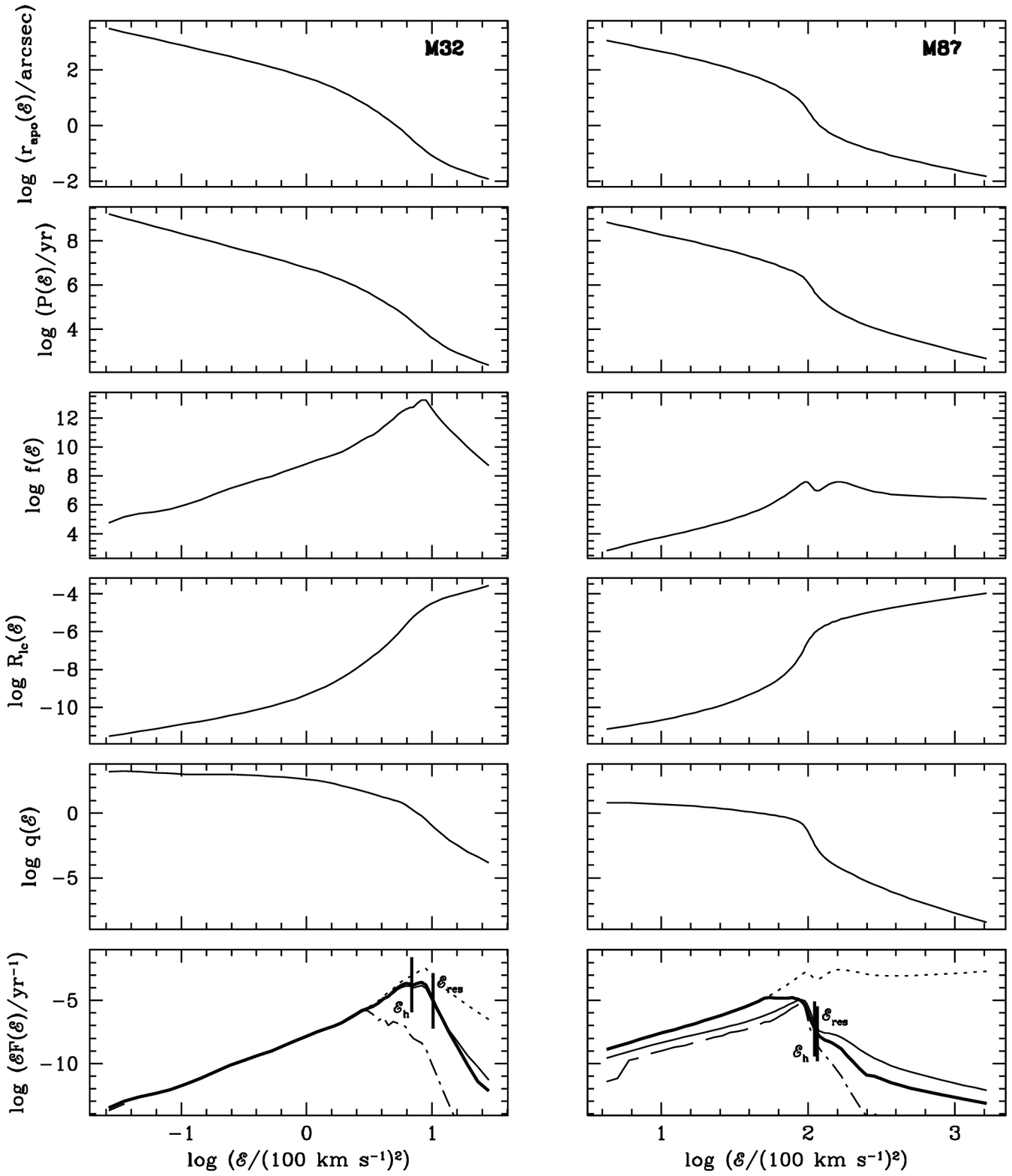,width=\hsize}}
\caption{{\bf Figure 2.}
Diffusion of stars into the loss cones of M32 (left) and M87 (right),
assuming that each is composed of stars of solar mass and radius.  The
top three panels on each side give the apocentre radius, orbital
period $P(\E)$ and DF $f(\E)$ as functions of energy~$\E$.  Below
these are the fraction $R\lc(\E)\equiv\jt^2(\E)/\jc^2(\E)$ of phase
space occupied by the loss cone at a given energy, and the parameter
$q(\E)$ that distinguishes between the ``diffusion'' $(q\ll-\ln R\lc)$
and ``pinhole'' $(q\gg-\ln R\lc)$ r\'egimes.
The bottom panels plot the flux of stars into the BH's maw due to: a
  full loss cone, $F^{\rm full}(\E)$ (Sec.~2 -- dotted curves);
  two-body relaxation of stars into the loss cone, $\Flcsol(\E)$
  (Sec.~3 -- light full curves);
   draining of the loss wedge,
   $\Fdrainsol(\E;10^{10}\yr)$ (Sec.~4 -- dot-dashed curves);
    diffusion of stars into the loss wedge, $\Flwsol(\E)$ (Sec.~4 --
  dashed curves).
  The heavy full curve plots our final flux, $\Fsol(\E)$, which is
  given by eq.~(51).
$\E_{\rm res}$~marks the energy above which
resonant relaxation becomes effective.
}
\endfigure

One process that has been neglected in this local Fokker--Planck
analysis is the resonant relaxation described by Rauch \& Tremaine
(1996) and Rauch \& Ingalls (1998): orbits in the nearly Keplerian
potential close to the BH can be thought of as slowly precessing
ellipses whose mutual torques allow much faster relaxation in angular
momentum.  This effect can enhance the relaxation rate a few-fold, but only
within a radius~$r_{\rm res}$ enclosing a mass $\sim0.1M_\bullet$ of
stars.  The corresponding energy $\E_{\rm res}$ is marked on Fig.~2.
Since $\Flc(\E_{\rm res})$ is relatively small, the enhancement due to
resonant relaxation will not result in a significant increase in the
total consumption rate.

Integrating over~$\E$, the total consumption rate for the edge-on
solar-composition model of M32 is $\Flcsol=7.6\times10^{-5}\yr^{-1}$.
For the more realistic mass function described in Appendix~A, this
translates to a flaring rate $\Flcmf=1.3\times10^{-4}\yr^{-1}$.  As
noted in Section~2, the DF (and therefore the flaring rate) depends on
the assumed inclination angle~$i$ of the galaxy.  The probability
$\pr(i\mid q')$ that a galaxy of observed flattening~$q'$ has
inclination~$i$ is given by equation~(6) of Paper~I.  Using this to
integrate the flaring rates for M32 over all~$i$ gives a mean rate
$\Flcmf=1.2\times10^{-4}\yr^{-1}$.  For M87, the ``consumption'' rate
$\Flcsol=2.1\times10^{-6}\yr^{-1}$, but, since only giant stars are
disrupted outside its BH's horizon, the more interesting visible
``flaring'' rate is a mere $\Flcmf=7.9\times10^{-9}\yr^{-1}$.

Table~1 lists both $\Flcsol$ and $\Flcmf$ for all galaxies in Paper~I
with best-fit $M_\bullet>0$.  Fig.~3 plots each as a function of bulge
luminosity~$L$.  The rate is controlled by the density of stars at
$\Eh$, so that compact galaxies with steep ``power-law'' central
density cusps (Lauer et al.\ 1995) have the highest flaring rates, of
up to about $10^{-4}\yr^{-1}$, while brighter, less dense ``core''
galaxies like M87 have lower rates of about $10^{-8}\yr^{-1}$. 
Our values for $\Flcsol$ are on average a factor of 4
larger than those calculated by Syer \& Ulmer (1999) using less
detailed models.

\begintable*{1}
\vbox to 646pt{\caption{{\bf Table 1.} }\vfil}
\endtable

\beginfigure*{3}
\centerline{
  \psfig{file=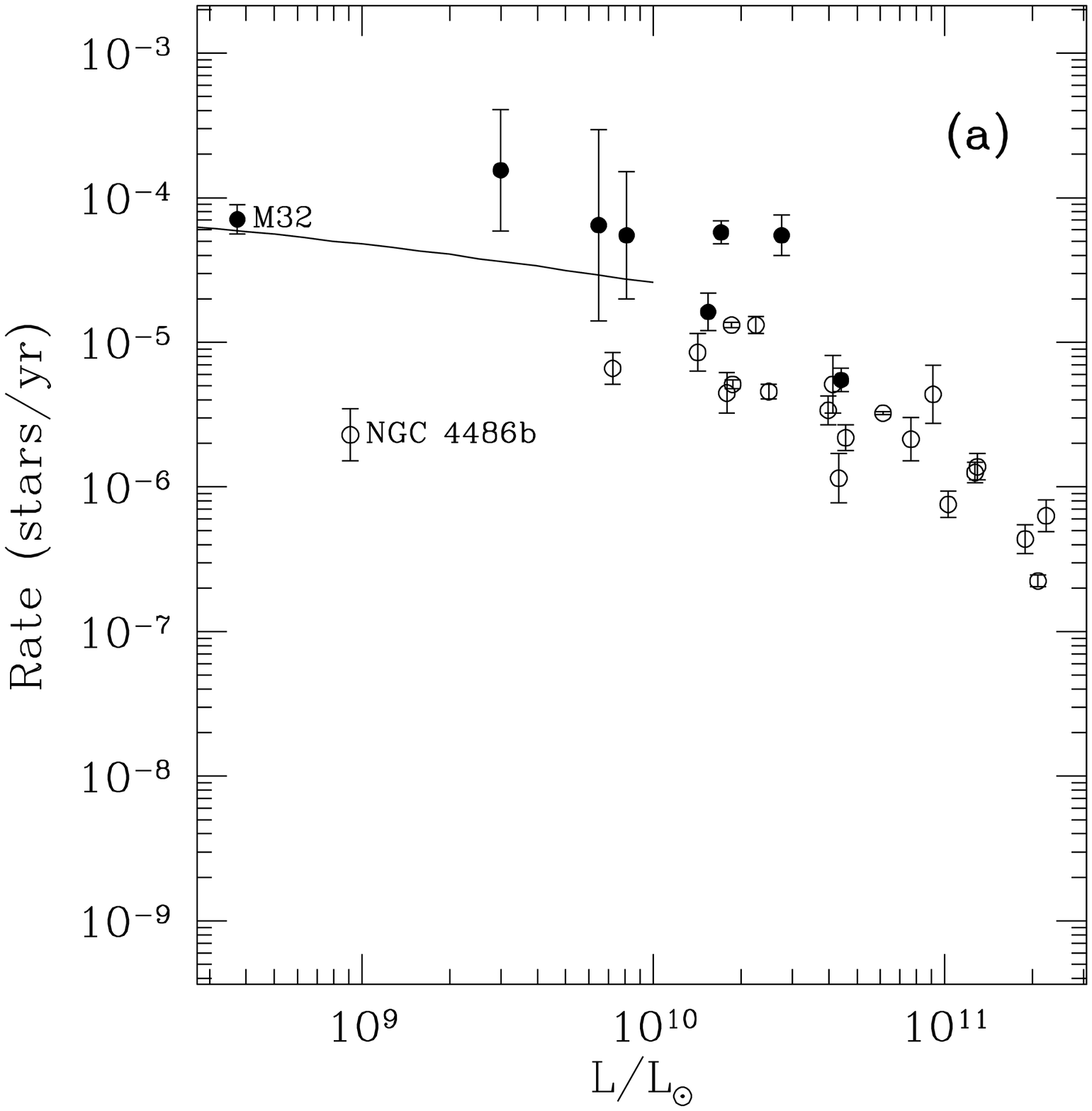,width=0.5\hsize}
  \psfig{file=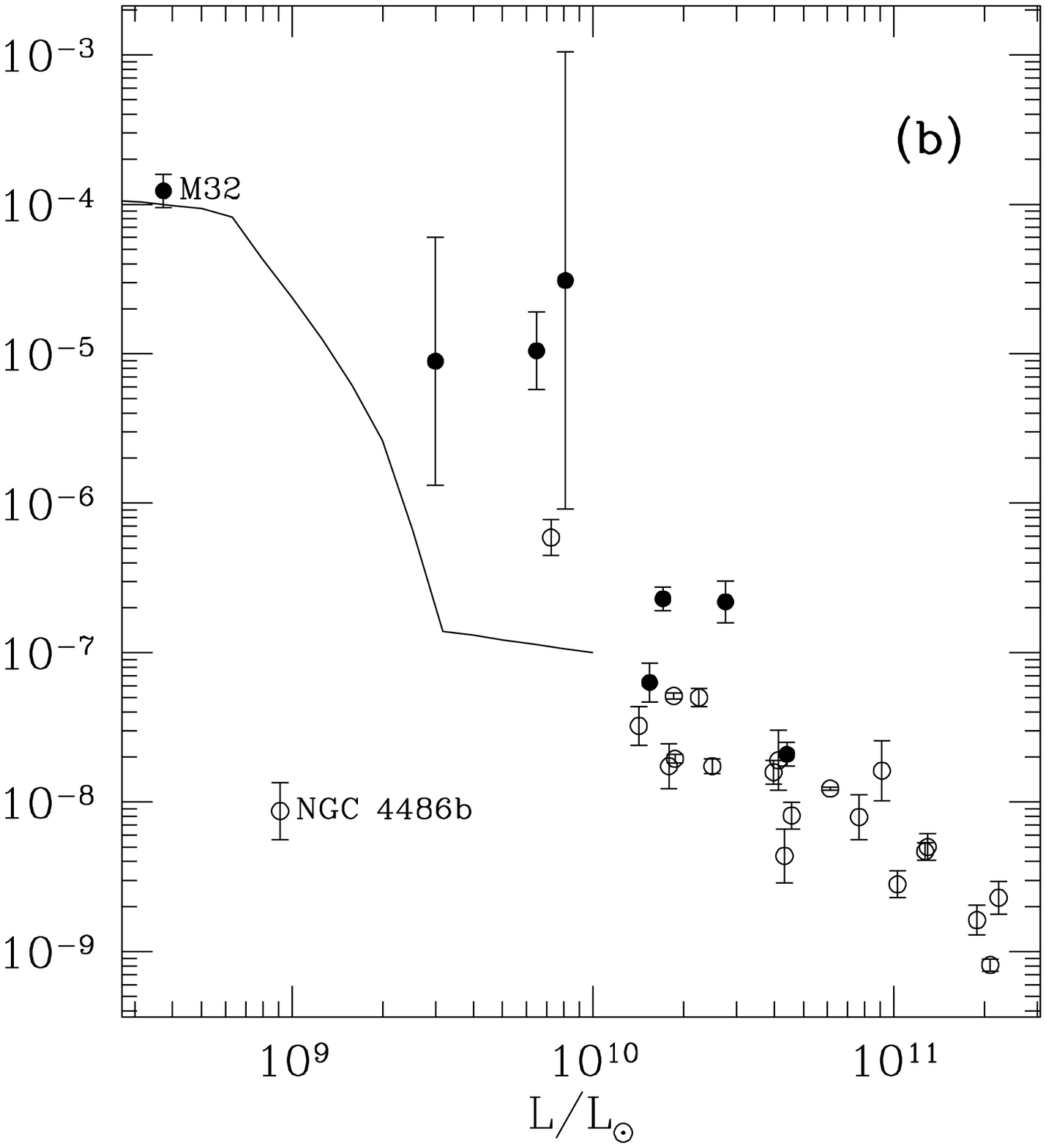,width=0.5\hsize}
}
\caption{{\bf Figure 3.}
Consumption rates due to two-body relaxation, plotted
as a function of bulge luminosity.  The panel on the left shows the
consumption rate~$\Flcsol$ (eq.~\ref{twobodynetrate}) assuming that each
galaxy is composed of stars of solar mass and radius.  The one on the
right shows the more interesting visible flaring rate~$\Flcmf$
assuming a more realistic mass function (Appendix~A).  Power-law and
core galaxies (Lauer et al.\ 1995) are plotted as filled and open
circles respectively.  The error bars give the uncertainty in the
rates due to the unknown inclination angle; the real uncertainties are
much larger.  The curves plot the rates predicted for the toy models
described in Section~6.}
\endfigure

\section {Effects of flattening: the loss wedge}

\subsection {Axisymmetry}

The results of the previous section are based on the assumption that
each galaxy is spherical, and therefore composed entirely of stars on
centrophobic loop orbits.  This is not the case, however, for more
realistic non-spherical galaxies.  Figs. 4(a) and~(b) show surfaces of
section obtained by following orbits with $J_z=0$, confined to the
plane $y=0$, in the potential of our axisymmetric edge-on model of
M32, and plotting $Y\equiv J_y/J\c$ versus colatitude $\theta$ at each
apocentre passage.  A significant portion of phase space in each case
is occupied by centrophilic orbits, which cross the $Y=0$ axis.  The
most tightly bound centrophilic orbits are regular ``lenses'', which
appear as bulls-eyes in Fig. 4(a) (Sridhar \& Touma~1997).  Further
out (Fig. 4(b)) the centrophilic orbits become stochastic (Gerhard \&
Binney~1985).

\beginfigure*{4}
\vskip0.3truecm
\centerline{
  \psfig{file=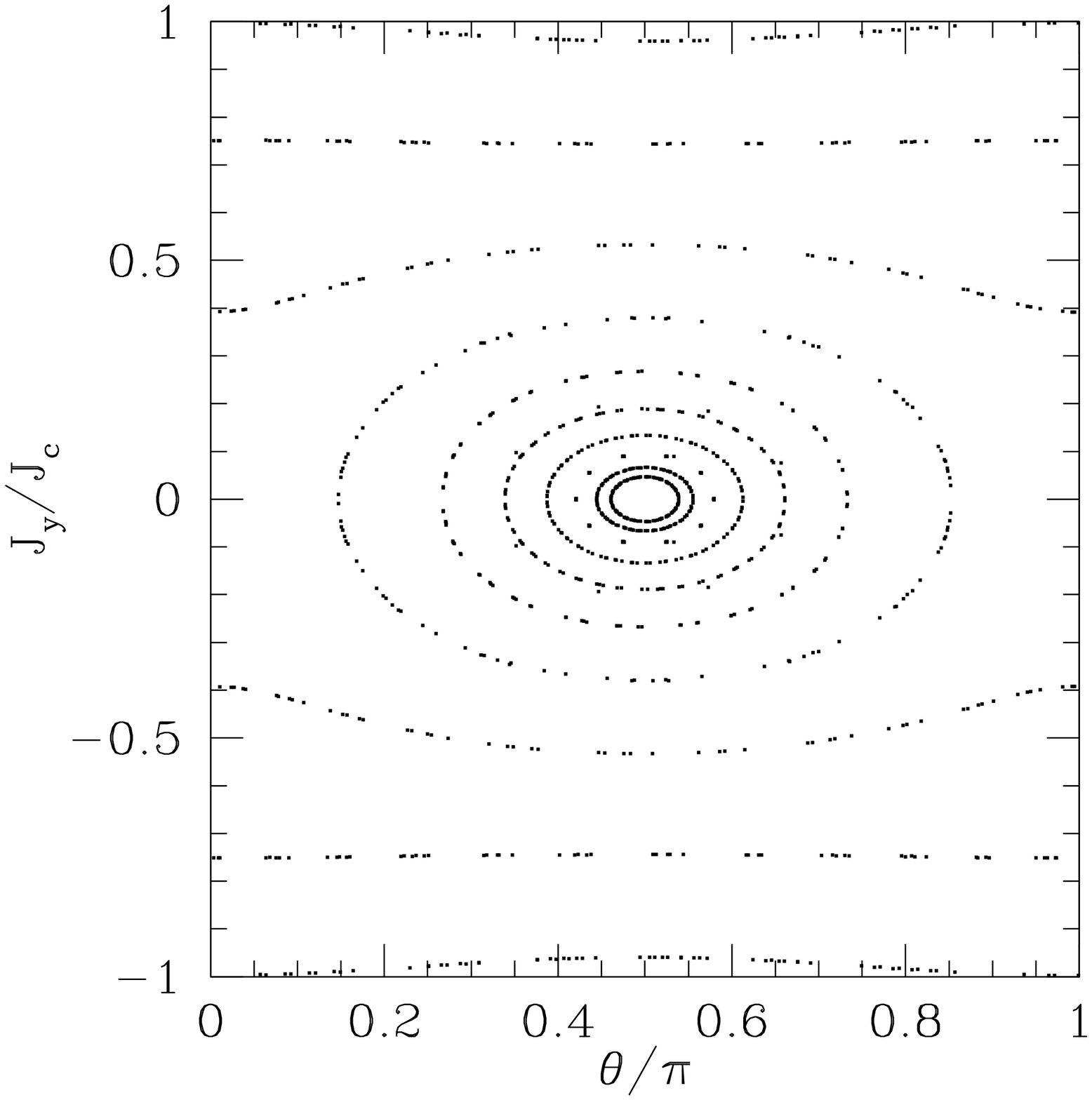,width=0.5\hsize}
  \psfig{file=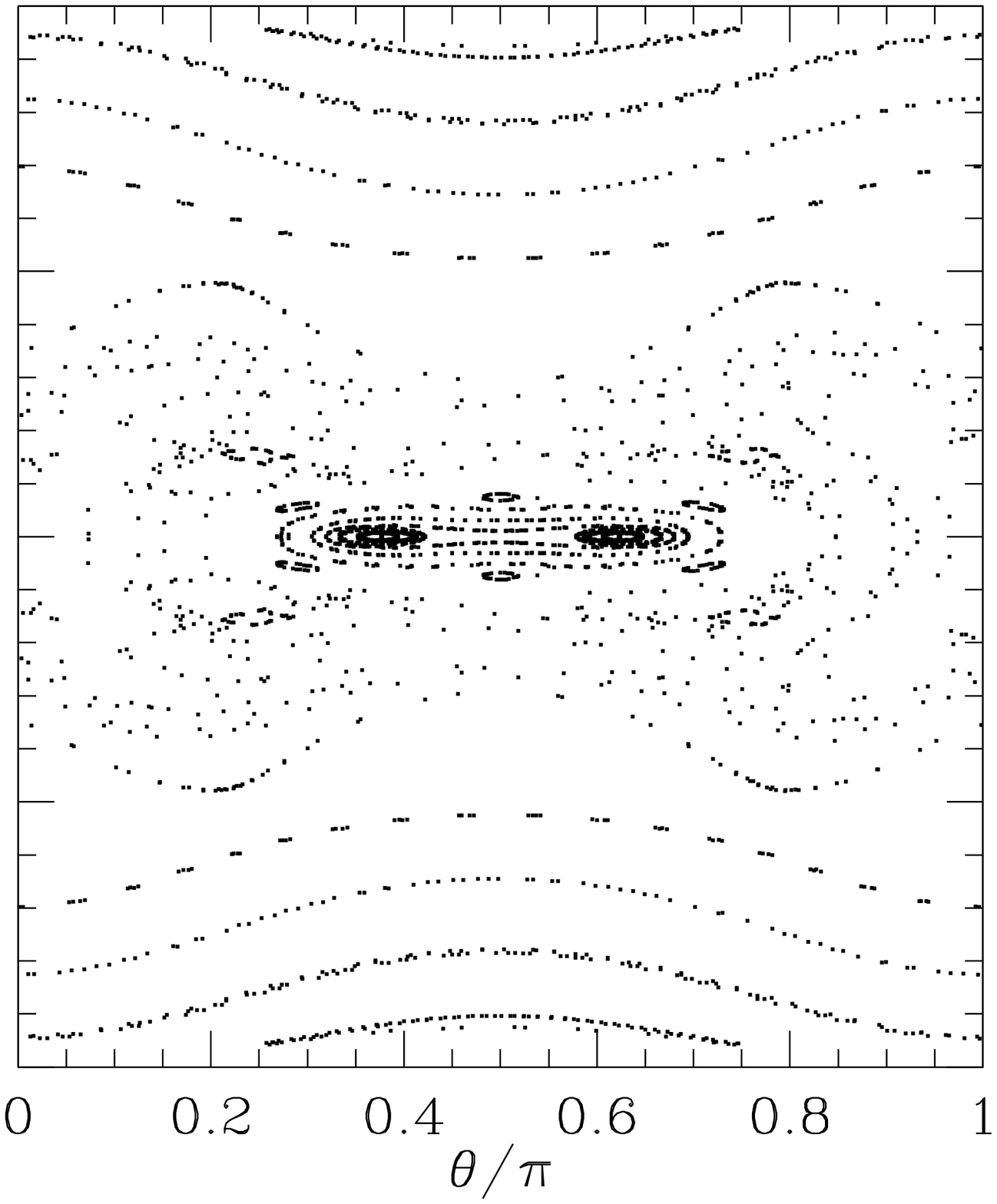,width=0.5\hsize}
}
\caption{{\bf Figure 4.}
Surfaces of section of $J_z=0$ orbits in our edge-on model of M32
obtained by plotting $J_y/J\c$ versus colatitute~$\theta$ at each
apocentre passage.  The panel on the left (right) plots orbits with
energies equal to that of a radial orbit with apocentre radius $1\pc$
($3\pc$). The bullseyes on the left are lens orbits (Sridhar \& Touma 1997).
}
\endfigure

The surfaces of section can be understood using the symplectic map of
Touma \& Tremaine (1997).  Values $(Y_n,\theta_n)$ of $(Y,\theta)$ at
successive apocentre passages are given by
$$\eqalign{
Y'_n & = Y_n-{1\over2}\epsilon\sin 2\theta_n\cr
\theta_{n+1} & = \theta_n + g(Y'_n)\cr
Y_{n+1} & = Y'_n -{1\over2}\epsilon\sin2\theta_{n+1}\cr
}\eqname\TTmap
$$
Here $\epsilon J\c$ is the
time-integral of the torque over one radial period, with 
$$\epsilon={4\over J\c}\int_{r_-}^{r_+} {\psi_2(r)\,\d r\over
  \left[2(\psi(r)-\E)-Y^2J\c^2/r^2\right]^{1/2}},
$$
where $\psi_2$ is the $\cos2\theta$ component of the potential,
and
$$g(Y)=2YJ\c\int_{r_-}^{r_+} {\d r\over r^2
  \left[2(\psi(r)-\E)-Y^2J\c^2/r^2\right]^{1/2}}
$$
is the advance in $\theta$ per radial period.  Touma \& Tremaine
demonstrate that this mapping provides a faithful representation of
orbit structure in scale-free potentials.  We find that it works just
as well for our models of real galaxies.  

These centrophilic orbits can have a dramatic effect on
the consumption rate, because stars with $J\gg J\lc$ can precess into the loss
cone so long as $|J_z|<J\lc$. Crudely speaking, the loss cone of the previous
section is replaced by a loss ``wedge'' with the same extent $|J_z|<J\lc$ in
the $J_z$ direction of phase space, but stretched in the $J$ direction to some
$J_l\gg J\lc$. 

To investigate this process, we work with the canonical
coordinate-momentum pairs $(J_z,\phi)$ and 
$(J_\theta,\theta)$, where $\theta$ and $\phi$ are the usual angles in
spherical coordinates and $J_\theta=r^2\dot\theta$. The total angular
momentum is given by
$$J^2 = {J_z^2\over\sin^2\theta} + J_\theta^2.
\eqname\totalj
$$
We are interested in the case $J_c\gg |J_\theta|\gg J\lc> |J_z|$, and we shall
assume that the variations in $J_\theta$ are described by the Touma-Tremaine
map \ref{TTmap} or approximations thereof. 
For $g(Y)\pmod\pi$ small, the mapping is well approximated by
motion under the averaged Hamiltonian
$$
H=G(Y)-{1\over2}\epsilon\cos2\theta
$$ 
where $Y=J_\theta/J_c$, $|Y|\ll1$ and 
$$
G(Y)\equiv\int_0^Y\left[g(Y')-g(0)\right]\,\d Y'.
$$
For most realistic galaxy models (e.g. central black hole plus star density
$\rho(r)\propto r^{-\gamma}$, with $\gamma<{5\over 2}$), $g(0)=2\pi$ and
$g'(0)$ is finite and negative. Thus we may write
$$
G(Y)\simeq -{1\over2}|g'(0)|Y^2,
$$
and the motion in the $(Y,\theta)$ plane is approximately that of a simple
pendulum:
$$Y^2(\theta)=Y_0^2-Y_{\rm m}^2\cos^2\theta.
\eqname\regorbity
$$
Here $Y_0$ is the peak angular momentum of the orbit, and $Y_{\rm m}$
is the peak angular momentum of the $H=-{1\over2}|\epsilon|$
separatrix orbit given by 
$$|G(Y_{\rm m})|=|\epsilon|,
$$
or, for small $\epsilon$, $Y_{\rm m}^2\simeq2|\epsilon|/|g'(0)|$.
Relativistic precession will cause the pericentre of an orbit with
semimajor axis $a$ and eccentricity $e$ to advance by an additional
amount 
$\Delta\theta\simeq3\pi r_{\rm S}/a(1-e^2)\simeq{3\over 2}\pi Y^2_{\rm
S}/Y^2$ per radial period, where $Y_{\rm S}J\c$ is  the angular
momentum of an orbit with pericentre at the Schwarzshild radius
$r_{\rm S}$.  This has negligible effect on our calculation of $Y_{\rm
m}$ using~\refeq1 since all our galaxies have $Y_{\rm m}\gg Y_{\rm S}$
(e.g., Fig.~4).

This averaged description of the motion does not work, however, below some
critical energy $\E_{\rm s}$ where the centrophilic
orbits become stochastic, as in Fig. 4(b).  $\E_{\rm s}$ can be estimated by
linearizing the map~\ref{TTmap} about $Y=0$ and making a trivial change of
variables, so that it becomes a symmetric version of the Chirikov--Taylor map
(see, e.g., Lichtenberg
\& Lieberman~1992 for extensive discussion).  The condition for
global stochasticity is (Touma \& Tremaine 1997, eq. 21)
$$|\epsilon|\gta{1\over2|g'(0)|}.
\eqname\stochcond
$$
We calculate the area of the stochastic region of
$(J_\theta,\theta)$ phase space by iterating the map~\ref{TTmap}
starting from $Y=0$ and a range of randomly chosen
$\theta\in [0,\pi]$.  The area
of phase space occupied by stochastic orbits is then taken to be 
$2\pi J_l$ where $J_l$ is twice the average value of $|J_\theta|$.  We
assume that orbits inside this stochastic region fill it uniformly.
Although this approach is quite crude (for example, the estimate of
the area can be affected by the presence of resonant ``boxlet'' orbits
described by Miralda--Escud\'e \& Schwarzschild~1989), it is more than
adequate for our purposes below.

\subsubsection{Draining a full loss wedge}
We first estimate the time required to drain the loss wedge, neglecting
refilling by two-body relaxation or other processes.  In the spirit of
our two-integral assumption, let us suppose that at time $T=0$ the
number of stars in a small interval of phase space is
$4\pi^2P(\E)f(\E)\d\E\,\d J_z\,\d J$ (cf. eq. \ref{nnnnn}), and that
the probability a given star is consumed in one radial period is $p$.
Then after time $T$ the loss wedge drains at a rate
$$\Fdrain(\E;T)\,\d\E=4\pi^2f(\E)\d\E\int \d J_z dJ\, p(1-p)^{T/P(\E)}.
\eqname\axirate
$$
After many orbits ($T\gg P(\E)$, $p\ll1$) this simplifies to
$$\Fdrain(\E;T)\,\d\E=4\pi^2f(\E)\d\E\int \d J_z dJ\, p\exp[-pT/P(\E)].
\eqname\axiratea
$$

We must now estimate the consumption probability $p$, equal to the fractional
time the star spends with total angular momentum $J<J\lc$. We
first consider the case of a star on a regular orbit (eq.~\ref{regorbity}).
Although these trajectories can be written explicitly in terms of elliptic
integrals, for the sake of simplicity we shall work with approximate orbits
valid for $Y_0\ll Y_{\rm m}$:
$$ J_\theta=J_0\sin\psi, \qquad \theta=\halff\pi+{J_0\over J_{\rm m}}\cos\psi,
$$
where $J_0=J_cY_0$, $J_{\rm m}=J_cY_{\rm m}$, and $\psi$ is a phase angle that
increases linearly with time. Therefore 
$$\eqalign{
p&\simeq {2\over\pi}\sin^{-1}\left({\sqrt{J\lc^2-J_z^2}\over J_0}\right)\cr
&\simeq{2\sqrt{J\lc^2-J_z^2}\over\pi J_0},\cr
}\eqname\regp
$$
for $J\lc\ll J_0<J_{\rm m}$ and $|J_z|<J\lc$, and zero
otherwise.

Next we estimate the area of $(J_\theta,\theta)$ phase space occupied
by orbits with peak angular momenta lying between $J_0$ and $J_0+\d
J_0$.  From~\ref{regorbity}, the area associated with lens orbits 
with peak angular momenta less than $J_0$ is $A(J_0)$, where
$A(J_0)\simeq\pi J_0^2/J_{\rm m}$ for $J_0\ll J_{\rm m}$ and $A(J_{\rm
m})=4J_{\rm m}$.  So, in equation~\ref{axiratea} we may replace the
area element $2\pi\d J$ by $\d A\simeq 2\pi J_0\d J_0/J_{\rm m}$ (for
$J_0\ll J_{\rm m}$). 

With these results, equation \ref{axiratea} becomes
$$\eqalign{\Fdrain(\E;T)\,\d\E&={4\pi^2\over J_{\rm m}}f(\E)\d\E\int \d J_z
J_0\d J_0
\,p\exp[-pT/P(\E)]\cr
&= 8\pi f(\E)\d\E\int_{-J\lc}^{J\lc} 
\d J_z \sqrt{J\lc^2-J_z^2}\cr
&\quad \times \int_1^\infty {\d x\over x^2}
\exp\left[-{2\sqrt{J\lc^2-J_z^2}\over\pi J_{\rm m}}{Tx\over P(\E)}\right],
\cr}
$$
where $x=J_{\rm m}/J_0$.  For times $T\ll P(\E)J_{\rm m}/J\lc$, the
exponential is roughly unity, and we have
$$
\Fdrain(\E;T)\,\d\E=4\pi^2J\lc^2f(\E)\d\E,
\eqname\mefull
$$
equal to the draining rate for a spherical system with full loss cone,
$F^{\rm full}(\E)$ (eq. \fulleq). For $T\gg P(\E)J_{\rm m}/J\lc$, the
draining rate is dominated by stars with $J\lc^2-J_z^2\ll J\lc^2$,
which give 
$$\Fdrain(\E;T)\,\d\E={\pi^4J_{\rm m}^3\over J\lc}\left[P(\E)\over T\right]^3
f(\E)\d\E;
\eqname\axiratereg
$$
the consumption rate declines as $T^{-3}$ (in the absence of refilling of the
loss wedge by two-body relaxation or other effects).

For stochastic ($\E<\E_{\rm s}$)
orbits we assume the orbits are uniformly distributed in the
$(J_\theta,\theta)$ plane for $|J_z|<J_l$. We then have from equation
\ref{totalj} that
$$p={1\over\pi J_l}\int\d\theta
\left(J\lc^2-{J_z^2\over\sin^2\theta}\right)^{1/2},
\eqname\tttyyy
$$
where the integral is taken over all $\theta\in[0,\pi]$ for which the
radicand is positive, and $p=0$ for centrophobic orbits with
$|J_\theta|>J_l$.  Then equation \ref{axiratea} yields
$$\Fdrain(\E;T)\,\d\E=4\pi^2f(\E)\d\E J_l\int_{-J\lc}^{J\lc} \d J_z\, 
p\exp[-pT/P(\E)].
$$
For times $T\ll P(\E)\jl/J\lc$, the exponential is roughly unity. 
Substituting equation \ref{tttyyy} and changing the order of
integration, we recover equation~\ref{mefull}. For $T\gg P(\E)\jl/J\lc$, the
draining rate is dominated by stars with $J\lc-|J_z|\ll J\lc$. 
In this case $p=(J\lc-|J_z|)/J_l$ and we find
$$\Fdrain(\E;T)\,\d\E=8\pi^2J_l^2\left[P(\E)\over T\right]^2f(\E)\d\E;
\eqname\axiratestoch
$$
the consumption rate declines as $T^{-2}$ and, remarkably, is independent of
$J\lc$; only the time required to reach this
asymptotic r\'egime depends on $J\lc$.

\subsubsection{Refilling the loss wedge by two-body relaxation}

These results neglect refilling of the loss wedge by two-body relaxation.  The
steady-state refilling rate can be calculated using a Fokker--Planck analysis
similar to that of the previous section, but considering diffusion in $J_z$
rather than in $R\equiv J^2/J\c^2$.  To zeroth order in~$J_z$, the frictional
diffusion coefficent $\langle\Delta J_z\rangle=r\sin\theta\langle\Delta
v_\phi\rangle=0$, while the diffusion coefficient
$$
D_z\equiv{1\over2}\left\langle(\Delta
J_z)^2\right\rangle={1\over2}r^2\sin^2\theta\left\langle(\Delta
v_\phi)^2\right\rangle\simeq{1\over4}\mu J\c^2\sin^2\theta,
$$
where $\mu$ is the diffusion coefficient of the previous section.  We
shall average $D_z$ over $\theta$ since in most galaxies the
precession time is short compared to the relaxation time; thus $\oa
D_z={1\over 8}\mu J_c^2$.

For regular $\E>\E_{\rm s}$ orbits the (orbit-averaged) flux of stars
through a surface of constant $J_z$ is
$$\eqalign{
{1\over2}\Flw(\E)\,\d\E 
   & = 2\pi \oa D_z P \int dJ_\theta d\theta {\p f\over\p J_z}\,\d\E\cr
   & = 2\pi q_z J\lc^2 A(J_{\rm m}) {\p f\over\p J_z}\,\d\E,\cr
}
\eqname\axifluxjz$$
where $q_z\equiv P\oa D_z/J\lc^2$ and we have assumed that $\p f/\p J_z$ is
approximately constant over the loss wedge.  In terms of $q\equiv
P\oa\mu/R\lc$ used in the previous section, $q_z=q/8$.  We include the factor
of ${1\over2}$ on the left-hand side of~\refeq1 because orbits with $J_z=\pm
J\lc$ contribute identical amounts to the total flux into the loss wedge.

We assume that the galaxy is in a steady state, at least for small
$|J_z|\ll J\c$.  Then $\Flw(\E,J_z)$ must be independent of $J_z$ for
$J\lc<|J_z|\ll J\c$. Thus 
$f(\E,J_z)=a(\E)+b(\E)|J_z|$. The rate of consumption
of stars is given by equation \ref{mefull} with $f(\E)=a(\E)$, and
this must equal $\Flw(\E)$; therefore $a(\E)=q_zA(J_{\rm
m})b(\E)/\pi$.
The corresponding ``isotropized'' DF
$$\bar f(\E)\equiv {b(\E)\over J\c}\int_0^{J\c}
\left({q_zA(J_{\rm m})\over\pi}+J_z\right)\,\d J_z,
$$
which is what one measures by inverting equation~\ref{eddington}.
Therefore, for $J_z\ll J\c$,
$$f(\E,J_z)=\bar f(\E){1+\pi |J_z|/(4q_zJ_{\rm m})
\over1+\pi J_c/(8q_zJ_{\rm m})}.
$$
Substituting into equation~\ref{axifluxjz}, the steady-state
consumption rate is
$$\Flw(\E)\,\d\E = 4\pi^2J\lc^2(\E){\iso f(\E)\,\d\E
\over 1+\pi J_c/(8q_zJ_{\rm m})}.
\eqname\axiflux
$$
The boundary between the ``diffusion'' and ``pinhole'' r\'e\-gimes for
the loss wedge occurs at 
$q_z\simeq 0.5J\c/J_{\rm m}$ (i.e., $q\simeq
4J\c/J_{\rm m}$), rather than at $q\simeq-\ln R\lc$ as in the case
when the galaxy is composed purely of centrophobic loop orbits.  In
the pinhole limit $q_z\gg J\c/J_{\rm m}$, the consumption rate
$\Flw(\E)$ is identical to $F^{\rm full}(\E)$: the
increase in the surface area of the loss wedge by a factor $\sim 4
J_{\rm m}/J\lc$ over that of the loss cone is exactly cancelled by the
decrease in the consumption probability per star per radial period.

For stochastic ($\E<\E_{\rm s}$) orbits, a similar calculation gives
$$\Flw(\E)\,\d\E=4\pi^2J\lc^2(\E){\iso f(\E)\,\d\E
\over 1+J\c/(4q_zJ_l)},
$$
so that the boundary between diffusion and pinhole r\'egimes occurs at
$q_z\simeq 0.25J\c/J_l$ (i.e., $q\simeq2J\c/J_l$). 
Equations \refeq1 and~\ref{axiflux} are identical when expressed in
terms of the area of $(J_\theta,\theta)$ phase space occupied by
centrophilic orbits ($2\pi J_l$ and 
$A(J_{\rm m})=4J_{\rm m}$ respectively).

The dashed curves in Fig.~2 show $\Flwsol(\E)$ for our edge-on models of M32
and M87.  The small differences between $\Flwsol(\E)$ and the corresponding
$\Flcsol(\E)$ are due solely to the differences between the steady-state
solutions for the two-dimensional loss-cone problem (in which
$f(J)\propto\ln(J^2/J_0^2)$) and the one-dimensional loss-wedge problem (where
$f(J_z)\propto a+b|J_z|$).  Taking all the galaxies in our sample, we find
that on average $\Flwsol$ is greater than $\Flcsol$ by a factor of 1.8,
although this ratio can be as large as 5 for the most most face-on, flattened
models in which a large portion of $(J_\theta,\theta)$ phase space is occupied
by centrophilic orbits.  The actual value of $\Flw$ itself typically only
increases by about 40\% from the edge-on model of a given galaxy to the most
face-on, flattened model, since the flatter model has fewer stars on
low-angular momentum orbits.

The dot-dashed curves in Fig.~2 show $\Fdrainsol(\E)$ assuming that
both galaxies have an ``age'' $T=10^{10}\yr$.  In moderately large
($L\gta10^{10}L_\odot$) galaxies, a significant fraction of giant
stars on centrophilic orbits may linger in the loss wedge for longer
than $10^{10}\yr$.  This fossil population 
can increase the flaring rates in these large
galaxies by orders of magnitude over $\Flwmf$.

Our final fuelling rate is
$$F(\E)=\max[\Flw(\E),\Fdrain(\E;10^{10}\yr)].
\eqname\finalflux
$$
For each of the galaxies in Paper~I with $M_\bullet>0$, Table~1 lists both the
``consumption'' rate $\Fsol$ assuming the galaxy is composed of stars of solar
mass and radius, and the visible flaring rate $\Fmf$ assuming the more
sensible stellar mass function of Appendix~A.  Figure~5 plots these rates as a
function of bulge luminosity.  Table~1 also lists the mean and standard
deviation of the distribution of the logarithm of the apocentre radii of the
disrupted stars; we see that most of the consumption is from radii $\gta 1$
arcsec, where the spatial distribution of stars is
well-determined by the observations.

\subsection {Triaxiality}

Similar arguments can be applied to 
triaxial galaxies.  Once
again, there will be some characteristic minimum angular momentum
$\js(\E)$ inside which most orbits are no longer loops, but centrophilic.
We suppose that at time $T=0$ the number of
stars on centrophilic orbits is $N(\E)\js^2/\jc^2\simeq 4\pi^2 P\js^2
f(\E)$.  Most of these orbits will be stochastic.  The probability
that a given star on such an orbit will be tidally consumed within a
radial orbital period is $\jt^2/\js^2$.  Then, after time~$T$ the
consumption rate is
$$
\Ftri(\E)\,\d\E = 4\pi^2f(\E)\jt^2(\E)
\exp\left[-{T\over P(\E)}{\jt^2(\E)\over\js^2(\E)}\right]\,\d\E.
\eqname\triaxialrate
$$
The refilling of the loss  region can be described using
equations similar to those in Section 4.1.2. 

These arguments assume that the potential of the galaxy remains fixed.  In
reality, the galaxy becomes slowly more axisymmetric from the centre
out, due to the action of the stochastic orbits.  Merritt \&
Quinlan~(1998) find that, for all but the most puny BHs, at any given
radius the galaxy becomes axisymmetric within $\sim10^2$ local orbital
periods.  So, after time $T$ we may take $J_{\rm s}(\E)=0$ for
$\E>\E_{\rm s}$, where $10^2P(\E_{\rm s})=T$.  For all our galaxies we
find that $\Ftri(\E)<F(\E)$, so that the additional effects of
triaxiality are negligible compared to those of axisymmetry.

\section {Contribution of consumed stars to $M_\bullet$}

The mass of stars
in a full loss wedge ranges from about $10^4M_\odot$ for small compact
galaxies (with $M_\bullet\sim10^6M_\odot$) up to about $10^7M_\odot$
for the largest galaxies (with $M_\bullet\sim10^{10}M_\odot$).  Thus
we can neglect the contribution of $\Fdrain$ to the growth of a
central BH, and estimate the contribution of consumed stars to
$M_\bullet$ by
$$M^{\rm eaten}\sim\Flwsol T M_\odot,
$$
where $T$ is the age of the galaxy.  Taking $T=10^{10}\yr$, Fig.~5(a)
shows that $M^{\rm eaten}\sim10^6M_\odot$, independent of galaxy
luminosity.  Only for the most compact galaxies (e.g., only for M32 in
our sample) has the consumption of stars had any significant effect on
the mass of the BH.  This result also implies that in most galaxies the
capture of stars is not strong enough to spin down the BH (e.g.,
Young~1977; Beloborodov et al.\ 1992).

\section {Detecting flares in surveys}

The sample of Paper I is strongly biased towards rare bright galaxies,
but nevertheless we can still use it to estimate the flaring rate
per unit volume.  Ferguson \& Sandage (1991) find that the luminosity
function of E+S0 galaxies in each of four nearby clusters is well
described by the Gaussian
$$N(L)\,\d\log L = {N_0\over\sqrt{2\pi}\Delta} \exp\left[
 -{1\over2}\left(\log L-\log L_0\over\Delta\right)^2\right]\d\log L,
\eqname\LF
$$
where $\Delta=0.6$, $L_0=1.3\times10^9h^{-2}L_\odot$ in the $B$
band and $h$ is the Hubble constant in units of $100\kms\,\hbox{Mpc}^{-1}$.
(The corresponding mean $V$-band luminosity $L_0=1.9\times10^9h^{-2}L_\odot$
assuming that $B-V=1$.)  We assume that this luminosity function also provides
a good description of field galaxies (Sandage, Tammann \& Yahil~1979).  We
estimate $N_0$ using the $B$-band luminosity function found by Efstathiou,
Ellis \& Peterson (1988), noting that, from their Table~4, approximately one
third of all galaxies brighter than about $10^9h^{-2}L_\odot$ are of type E or
S0.  This yields $N_0=8.3\times10^{-3}h^3\Mpc^{-3}$.  The corresponding mean
$B$-band luminosity density of field E+S0 galaxies is then
$2.8\times10^7h\,L_\odot\Mpc^{-3}$.  For comparison, the luminosity density of
dE+E+S0 galaxies is $4.4\times10^7h\,L_\odot\Mpc^{-3}$ (Yahil, Sandage \&
Tammann~1980) and the luminosity density of all hot components (E+S0 galaxies
and spiral bulges) is $5.4\times10^7h\,L_\odot\Mpc^{-3}$, estimated using
Efstathiou et al.'s (1988) mean galaxy luminosity density $j=1.8\times10^8h\,
L_\odot\Mpc^{-3}$ and Schechter \& Dressler's (1987) result that bulges
contribute approximately 30\% to this (see also Appendix~B of Faber et al.\
1997).

Figs. 3 and~5 hint that the consumption rates are roughly constant functions
of~$L$.  We now show that this is indeed the case by using known correlations
of galaxy properties to construct a sequence of toy galaxy models and
calculating their consumption rates.  Following Paper~I, we assume that every
bulge has a BH of mass
$$
M_\bullet =
6.1\times10^8h^{-1}\left(L\over10^{10}h^{-2}L_\odot\right)^{1.17}M_\odot,
\eqname\MLcorrel
$$
and a stellar mass-to-light ratio
$$\Upsilon_V =
6.6h\left(L\over10^{10}h^{-2}L_\odot\right)^{0.18}\Upsilon_\odot.
$$
Galaxies with $L\lta2\times10^9h^{-2}L_\odot$ ($M_\bullet\lta10^8M_\odot$)
will dominate the consumption rate (Fig.~5).  These galaxies have
steep power-law central density cusps (Faber et al.~1997), which we
approximate using an E2 Jaffe (1983) model.  For the effective radius
of each model we take
$$R_{\rm eff}=2.4h^{-1}\left(L\over10^{10}h^{-2}L_\odot\right)^{0.65}\kpc,
$$
obtained by fitting to the galaxies in Faber et al. (1997).
We find that 
$$\Fsol(L)=2.7\times10^{-5}h^{2/3}
\left(L\over10^{10}h^{-2}L_\odot\right)^{-0.22}\yr^{-1},
$$
almost independent of luminosity.  We plot this consumption rate
$\Fsol(L)$ and the flaring rate $\Fmf(L)$ for these toy models on
Fig.~5 (and also $\Flcsol(L)$ and $\Flcmf(L)$ on Fig.~3).  They are
broadly consistent with the rates calculated for real galaxies, given
the small number of the latter with $L\lta10^{10}L_\odot$, and our
neglect of the substantial scatter in the relation between $M_\bullet$
and~$L$.  Integrating the toy models' $\Fmf(L)$ weighted by the
luminosity function~\ref{LF} gives a flaring rate per unit volume of
$7.5\times10^{-7}h^{11/3}\yr^{-1}\Mpc^{-3}$ for
E+S0 galaxies.

There is currently little evidence for or against the presence of BHs
in the centres of late-type bulge-less spiral galaxies, but BHs do
appear to be common features in bulges of type Sbc and earlier
(Richstone et al.\ 1998).  These early-type spirals have approximately
the same luminosity distribution~\ref{LF} as E+S0 galaxies, and are
about as common (Binggeli, Sandage \& Tammann~1988; see also above).
Neither of the two spirals
(M31 and NGC~4594) in Paper~I show any peculiarities in their
consumption rates compared to the other E+S0 galaxies.  Therefore we
multiply the flaring rate for E+S0 galaxies by 2 to obtain a total
flaring rate per unit volume of
$1.5\times10^{-6}h^{11/3}\yr^{-1}\Mpc^{-3}$.

There are a couple of major uncertainties in this rate, over and above
those we discuss in the next section.  First, the evidence for our
assumed $M_\bullet$--$L$ correlation~\ref{MLcorrel} is slight for
$L\lta10^{10}L_\odot$.  Second, the density profiles of real galaxies
may not follow the steep $\nu\sim r^{-2}$ profile of our assumed Jaffe
models close to their centres.  For example, the density profile of
M32 is observed to turn over at small radii (Section~2).  Modelling it
with a Jaffe profile would overestimate $\Flw$ by about a factor of~2.

\beginfigure*{5}
\centerline{
 \psfig{file=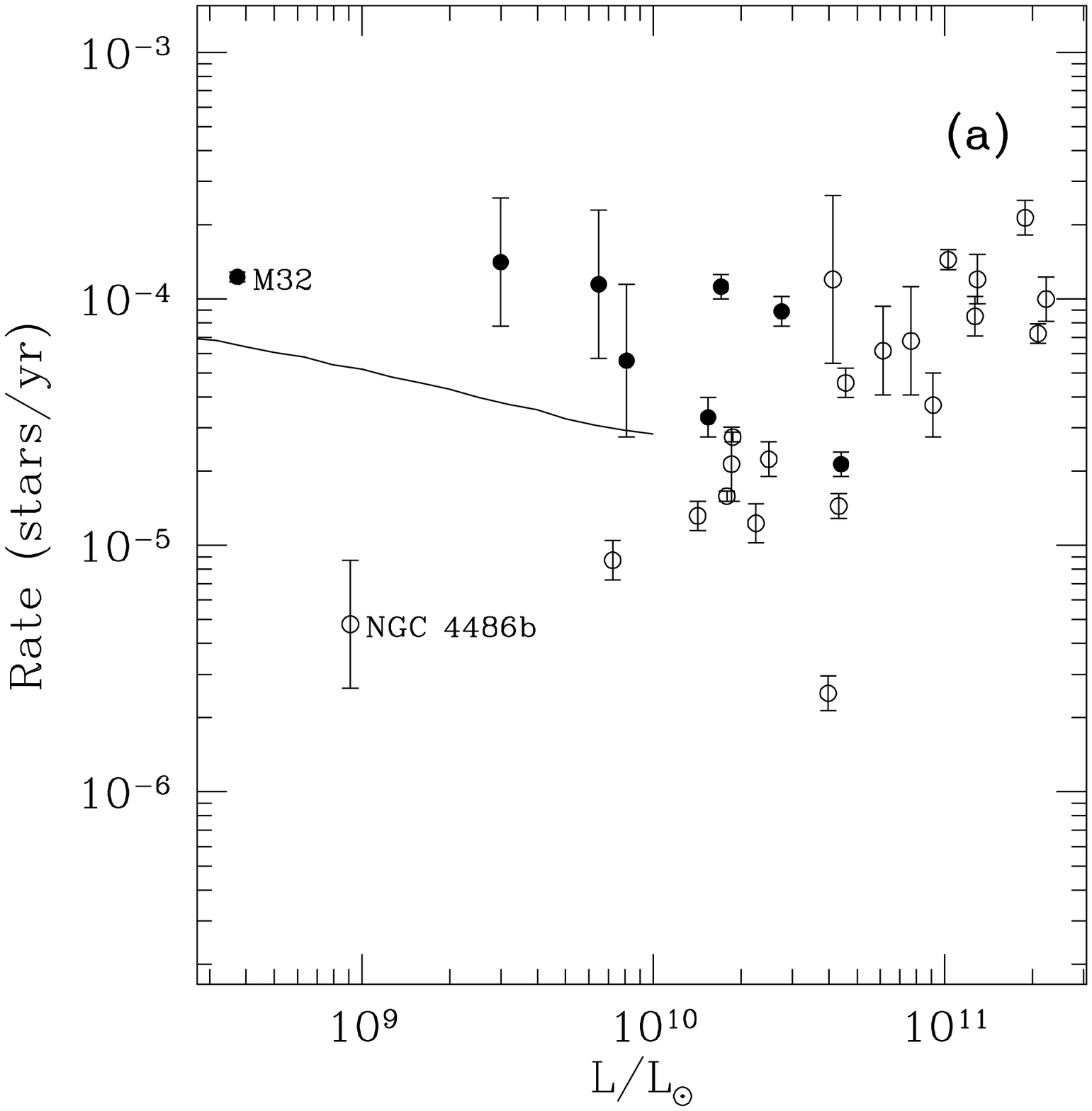,width=0.5\hsize}
 \psfig{file=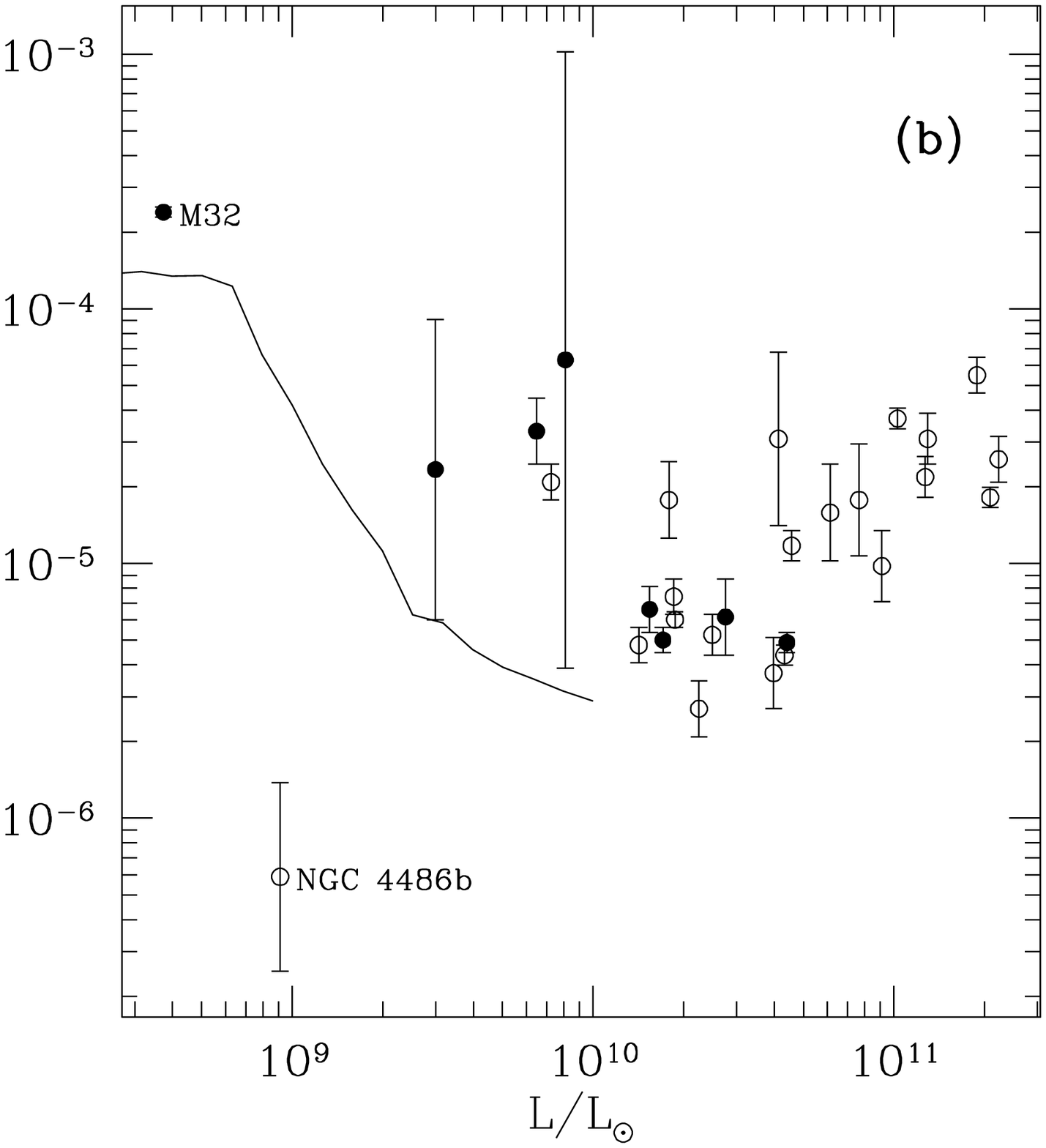,width=0.5\hsize}
}
\caption{{\bf Figure 5.}
As for Fig.~3, but plotting total consumption rates
(eq.~\ref{finalflux}) taking account of flattening and two-body
relaxation.  The panel on the left shows the consumption rate~$\Fsol$
assuming that each galaxy is composed of stars of solar mass and
radius.  The one on the right shows the visible flaring rate~$\Fmf$
assuming the more realistic mass function in Appendix~A.  Power-law
and core galaxies (Lauer et al.\ 1995) are plotted as filled and open
circles respectively.  The error bars give the uncertainty in the
rates due to the unknown inclination angle; the real uncertainties are
larger.  The curves plot the rates for the toy models described in
Section~6.}
\endfigure

According to the thick disk models of Ulmer (1998), disruption of a
solar-type star by a $10^6M_\odot$ ($10^7M_\odot$) BH should create a
flare with $V$-band luminosity of at least $2\times10^7L_\odot$
($8\times10^8L_\odot$).  From the correlation~\refeq1, these BH masses
correspond to host bulge $V$-band luminosities of $3\times10^8L_\odot$
($1.3\times10^9L_\odot$) -- thus the flare luminosity is about 10\% to 60\%
of the luminosity of the surrounding bulge.  In the $U$-band the
flares outshine the galaxy by at least a factor of four.  Thus, rather
than monitoring the nearest $10^5$ or so closest candidate galaxies
for evidence of flares, it would be much more economical to survey
small areas of the sky deeply, as in done in current supernovae
searches (e.g., Pain et al.\ 1996).  For a survey sensitive to all
flares within a redshift~$z$, the expected detection rate is
approximately 
$$0.11h^{2/3}\left(z\over0.3\right)^3
\hbox{per square degree per year}.
$$
with an uncertainty of at least a factor of 2.  This flaring rate is
about 0.01 times the Type Ia supernova rate.

In their $R$-band survey, Pain et al.\ (1996) can
just detect a supernova of apparent magnitude 21.8 in the centre of a
galaxy of magnitude 20.8 (their Fig.~1).  Thus the brightest flares,
yielding a 40\% increase in the luminosity of a
$L_V=2\times10^9L_\odot$ galaxy, are just detectable out to $z=0.3$,
and an $R$-band survey will not yield more than about one flare per
ten square degrees per year.  It might be possible to carry out
significantly deeper searches in the $V$-band (corresponding to the
flares' rest-frame $U$ band).

These results imply that a flare with $L\sim 10^{7.5}L_\odot$ occurred
in the centre of M32 within the last $10^4$ yr or so. At present there
is no evidence for non-thermal emission from the centre of M32: HST
photometry shows no significant evidence for a colour gradient in the
central pixel (Lauer et al. 1998), setting an upper limit of $\sim
10^{4.5}L_\odot$ to any non-thermal source. Thus the flare luminosity
must decay by at least a factor of $10^3$ over $10^4$ yr.

\section{Possible shortcomings of the models}

The models above are based on some quite strong assumptions.  Here we discuss
a number of ways in which real galaxies might differ from the models.

{\it Anisotropy:} Our two-integral assumption is most reasonable for power-law
galaxies, since their rapid rotation limits the amount of radial anisotropy
they may have.  No such constraint exists for larger core galaxies which tend
to rotate slowly, if at all.  Any increase in the number of stars on very
low-$J$ orbits will obviously enhance the flaring rates. Recent models
(Merritt \& Oh 1997; Rix et al.\ 1997; Gerhard et al.\ 1998) suggest that core
galaxies could have $\sigma_r^2/\sigma_\phi^2$ lying between 1.2 and 1.4
(under the assumption of spherical symmetry). However, even detailed
kinematic modelling of individual galaxies can only yield information on
their DFs for $J\gta0.1J\c\gg J\lc$, and thus does not greatly improve the
reliability of estimates of flaring rates. 

{\it Tidal capture:} The calculations of Sections 3 and~4 give rates
for the {\it direct} disruption of stars.  They neglect the
possibility that stars that come within a few times $r_{\rm t}$ of the
BH may be tidally captured by the BH and then disrupted during a
subsequent pericentre passage (e.g., Frank \& Rees~1976; Novikov,
Pethick \& Polnarev~1992; Diener et al.\ 1995).  The radius for tidal
capture is at most $\sim 3r_{\rm t}$ (Novikov et al.\ 1992), and thus tidal
capture could enhance the disruption rates by up to a factor
three.  Similar arguments apply to any dense extended disc (e.g.,
Norman \& Silk~1983) that may surround the BH.

{\it BH wandering:} A BH can wander from the centre of the galaxy as a
result of Poisson noise from the stellar distribution (Bahcall \&
Wolf~1976), weakly damped $l=1$ lop-sided modes of oscillation
(Weinberg~1994), an eccentric stellar disk (Tremaine 1995), or other
processes; all of these are poorly understood.  It is unclear what
effect wandering would have on the fuelling rate.  
If the wandering timescale is long
compared to stars' orbital timescales then clearly the resulting
refilling of the loss wedge would result in an increase in the
disruption rates.  On the other hand, if the wandering timescale is
much shorter than the orbital timescale and the BH has a typical
displacement $r_0\ll r_{\rm h}$ from the centre of the stellar
distribution, then stars with pericentres within $r_0$ (rather than
$r_{\rm t}$) can potentially be consumed.  From
equation~\ref{LCboundary}, this is a factor $\sim r_0/r_{\rm t}$ more
stars than if the BH stayed at the centre of the galaxy.  But the
probability that any given one of these stars will be consumed is only
$\sim r_{\rm t}^2/r_0^2$, so that there would be a net {\it decrease}
of $\sim r_{\rm t}/r_0$ in the consumption rate if the BH wandering
were fast enough.

{\it Other relativistic effects:} Our calculations use $\eta=0.844$ in
equation~\ref{tidalrad} for $r_{\rm t}$.  This value is based on
Diener et al.'s (1995) Newtonian calculations, which strictly apply only for
$M_\bullet\ll10^8M_\odot$.  We have also assumed that stars for which
$r_{\rm t}\gta r_{\rm S}$ are consumed whole by the BH, without a
visible flare.  This only applies to Schwarzschild BHs.  For Kerr BHs,
Beloborodov et al.\ (1992) show that $r_{\rm t}$ depends on the
direction at which the star approaches the BH, with the angle-averaged
$r_{\rm t}$ approximately equal to the $r_{\rm t}$ obtained assuming a
Schwarzschild BH.  Thus a spinning BH of mass $M_\bullet>10^8M_\odot$
could disrupt a main-sequence star that approaches from a favourable
direction.  

\section{Conclusions}

We have calculated the rates of
disruption of stars by central BHs in two-integral models of nearby
galaxies.  Our simplest calculation (Section~3) is based on the
assumption that all stars are on centrophobic loop orbits.  Then there
is a small portion of phase space, the loss cone $J<J\lc$, within
which stars will be consumed by the BH within an orbital period.
Two-body relaxation causes stars to diffuse into this loss cone, but
at a very low rate: only for loosely bound orbits in the faintest,
most compact galaxies is the diffusion strong enough to keep the loss
cone full.

Real galaxies are not, however, composed entirely of stars on loop
orbits.  In the presence of the slightest degree of flattening, a
significant portion of phase space is occupied by centrophilic orbits,
so that the loss cone becomes a ``loss wedge'' (Section 4). 
In a steady state, this loss wedge feeds
stars to the BH at a rate that is almost always faster than would
be obtained if all stars were on centrophilic loop orbits; however,
the enhancement is typically less than a factor 2 or so.  There is
nevertheless one important difference in the nature of the disruption
in the two cases: stars on centrophobic orbits are more likely to
approach the BH on deeply plunging radial orbits, with pericentres
well within~$r_{\rm t}$.  Stars on these ultraclose orbits are
expected to ``pancake'', possibly with an explosive release of energy
(e.g., Carter \& Luminet~1982; Rees~1988, 1998).

The best places to look for evidence of tidal disruption are faint
($L_V\lta10^{10}L_\odot$) bulges with steep, power-law central density
cusps.  In such galaxies, two-body relaxation causes a main-sequence
star to wander into the BH's maw every $10^4$ to $10^5\yr$, yielding a
visible flare. Surveys sensitive to flares out to a redshift $z=0.3$
(e.g., current $R$-band supernovae searches) are expected to see only
about one flare per 10 square degrees per year.  It might be possible
that $V$-band surveys could yield much higher rates. 
Apart from the extra complications introduced by centrophilic
orbits mentioned above, our results are in broad agreement with those
of Syer \& Ulmer (1999).

Flares occur much less frequently in larger galaxies (assuming that
their DFs are roughly isotropic), partly because such galaxies have
BHs with $M_\bullet\gta10^8M_\odot$ that swallow main-sequence stars
whole, and partly because such galaxies are less centrally
concentrated, meaning that their timescales are much longer.  The
flaring rate from the two-body relaxation of giants into a large
galaxy's loss cone can be as low as $10^{-9}\yr^{-1}$.  On the other
hand, the slow timescales mean that the depletion of the stock of
stars on centrophilic orbits occurs much more slowly than in compact
power-law galaxies.  We estimate that the disruption of giant stars on
low-$J$ centrophilic orbits could occur as often as once every
$10^5\yr$ or so in the largest galaxies.  However, very large galaxies
are extremely rare, and so will not affect the expected detection rate
in surveys.   This enhanced rate of fuelling from
centrophilic orbits is also important in other contexts (e.g.,
hardening of black-hole binaries -- Begelman, Blandford \& Rees\ 1980;
generation of gravitational waves from stellar remnants passing close
to the BH -- Sigurdsson \& Rees\ 1997).

At least two processes might change the flaring rate dramatically, neither of
which unfortunately is easy to model.  First, the BH may wander from the
centre of mass of the galaxy. Second, a modest degree of radial anisotropy
could increase the flaring rate dramatically, although this enhancement is
very sensitive to the observationally inaccessible details of the DF for
near-radial orbits.

\section*{Acknowledgments}

We thank David Merritt, Saul Perlmutter, Martin Rees, Dave Syer and our
collaborators on the Nuker team for useful conversations and correspondence.
Financial support to JM was provided by NSERC and PPARC. Partial support for
this work was provided by NASA through grant number NAG5-7066, and grant
GO-07388.06-96A from the Space Telescope Science Institute, which is operated
by the Association of Universities for Research in Astronomy, Inc., under NASA
contract NAS5-26555.

\def\apj #1 #2{ApJ, #1, #2}
\def\apjs #1 #2{ApJS, #1, #2}
\def\aj #1 #2{AJ, #1, #2}
\def\mn #1 #2{MNRAS, #1, #2}
\def\aa #1 #2{A\&A, #1, #2}
\def\araa #1 #2{ARA\&A, #1, #2}

\appendix

\section {Generalization to a spectrum of stellar masses}

{\bf Main-sequence stars:}\qquad It is straightforward to generalize the
calculations in the body of this paper to a range of stellar masses
and radii.  Let us assume that the probability of finding a main-sequence star
with mass in the range $m_\star$ to $m_\star+\d m_\star$ within some
phase-space volume $\d^3\b x\,\d^3\b v$ around $(\b x,\b v)$ is
$$f(\b x,\b v;m_\star)\,\d^3\b x\,\d^3\b v\,\d m_\star 
 = f_\odot(\b x,\b v)n(m_\star)\,\d^3\b x\,\d^3\b v\,\d m_\star
$$
with
$$n(m_\star)=\cases{A\left(m_\star/ M_\odot\right)^{-x}, 
                                     & $m_1<m_\star<m_2$  \cr
                    0                & otherwise, \cr
}
$$
and $f_\odot$ the DF obtained by inverting equation~\ref{eddington}
assuming the galaxy is composed of solar-type stars.  We take
$m_1=0.08M_\odot$, $m_2=1M_\odot$ (no recent star formation) and
$x=2.35$ (Salpeter mass function).
This DF has to reproduce the mass distribution and mass-to-light ratio 
inferred from
observations, so that (see also equation~\ref{eddington})
$$\eqalign{
\rho(0,z)&=
4\pi\int_0^{\psi(0,z)}f_\odot(\E,0)\sqrt{2[\psi(0,z)-\E]}\,\d\E\times\cr
&\quad\qquad\int_{m_1}^{m_2}m_\star n(m_\star)\,\d m_\star.\cr
}$$
Therefore $\int_{m_1}^{m_2}m_\star n(m_\star)\,\d m_\star=M_\odot$ so that 
$$A={2-x\over M_\odot}\left[\left(m_2\over M_\odot\right)^{2-x}-
                      \left(m_1\over M_\odot\right)^{2-x}\right]^{-1}
=0.246M_\odot^{-1}.
$$

For lower main-sequence stars, $r_\star\sim m_\star^{0.8}$ (Kippenhahn
\& Weigert~1990).  So from equation~\ref{tidalrad}, the tidal
disruption radius $r_{\rm t}\propto m_\star^{0.467}$.  A main-sequence
star whose mass satisfies 
$$m_\star<m_{\rm min}(M_\bullet)\equiv
a\left(M_\bullet\over10^8M_\odot\right)^{1.43}M_\odot,
$$
will be swallowed whole by the BH; here $a$ is a constant of order unity
(=1.05 for $n=3$ polytropes; see Diener et al. 1995). 

{}From Appendix B, the diffusion coefficient for two-body 
relaxation
$$\mu_{\rm MF}=\mu_\odot\int_{m_1}^{m_2}
\left(m_\star\over M_\odot\right)^2n(m_\star)\d m_\star
\simeq0.31\mu_\odot.
$$
If we ignore the $\ln R_0$ factor in equation~\ref{twobodynetrate},
two-body relaxation gives a flaring rate 
$$\Flcmf(\E)=\Flcsol(\E){\mu_{\rm MF}\over\mu_\odot}
\int_{\max[m_1,m_{\rm min}(M_\bullet)]}^{m_2}
n(m_\star)\d m_\star,
\eqname\twobodyscale
$$
which for $M_\bullet<1.7\times10^7M_\odot$ is $1.66\Flcsol$. 
In calculating the $\Flcmf$ reported in Table~1, we have
taken into account the $\ln R_0$ factor in
equation~\ref{twobodynetrate}, giving slightly higher results than
this.  The effects of the mass spectrum on the $\Flw$ given
by~\ref{axiflux}, and on the $\Fdrain$ given by equations
\ref{axiratereg} and~\ref{axiratestoch}, are dealt with in a similar
way.

{\bf Giant stars:}\qquad A small fraction $g\ll1$ of stars will be
giants with some characteristic radius $r_{\rm g}$.  These are the
only stars that produce flares in large galaxies with
$M_\bullet>10^8M_\bullet$.  We take $g=0.01$ and $r_{\rm
g}=15r_\odot$.  From equations \ref{twobodynetrate} and~\ref{axiflux},
giants contribute a factor $g(\mu_{\rm MF}/\mu_\odot)$ times $\Flcsol$
or $\Flwsol$ to the rates in the diffusion limit, rising to a factor
$g(r_{\rm g}/r_\odot)(\mu_{\rm MF}/\mu_\odot)$ in the pinhole limit.
It is straightforward to calculate their contribution to $\Fdrainmf$
using equations \ref{axiratereg} and~\ref{axiratestoch}.

Finally, there is a further way of feeding giant stars to the BH.  In
the normal course of stellar evolution, a main-sequence star on a
low angular-momentum orbit may turn into a giant and suddenly find
itself in the wider, giant loss cone (Syer \& Ulmer~1999).  We ignore
this process, as it has negligible effect compared to, for example,
the draining of giant stars on the loss wedge (Section 4.1.1).

\section {Diffusion coefficients}

In this Appendix we derive expressions for the diffusion coefficients
used in equation~\ref{FP} in the limit $R\rightarrow0$.  Because of
the presence of the loss cone, the steady-state distribution of
scatterers is not quite isotropic.  It is, however, reasonable to
calculate the diffusion coefficients using the isotropized
distribution function
$$\bar f(\E)\equiv \int_0^1 f(\E,R)\,\d R.
$$

We make the reasonable approximation that all encounters take place
instantaneously and so change the scattered star's velocity but not
its position.  In addition, we make the usual (though more dubious)
assumption that the distribution of scatterers is homogeneous in
space.  Since $R\equiv r^2v_{\rm t}^2/J^2\c(\E)$, where $v_{\rm
t}^2=v_\phi^2+v_\theta^2$, we can immediately use equation (8--64) of
Binney \& Tremaine (1987) to show that
$$\DR={32\pi^2r^2G^2m_\star^2\ln\Lambda\over3J\c^2} \left(
3I_{1\over2}-I_{3\over2}+2I_0 \right)+{\cal O}(R),$$
where
$$
\eqalign{%
I_0 & \equiv \int_0^\E \iso f(\E')\,\d\E',\cr
I_{n\over2} & \equiv 
\left[2\left(\psi(r)-\E\right)\right]^{-{n\over2}}
  \int_\E^{\psi(r)}
  \left[2\left(\psi(r)-\E'\right)\right]^{n\over2}\iso f(\E') \,\d\E',\cr}
$$
and $\ln\Lambda$ is the usual Coloumb logarithm.  We follow Spitzer \&
Hart (1971) and take $\Lambda=0.4M_\bullet/m_\star$.

The second-order diffusion coefficient is $\DRR={r^4\left(\Delta
v_t^2\right)^2/ J\c^4(\E)}$.  Since 
$$\eqalign{
\left(\Delta v_t^2\right)^2 & =
4v_\theta^2(\Delta v_\theta)^2 + 4v_\phi^2(\Delta v_\phi)^2 + \cr
&\qquad 8v_\theta v_\phi\Delta v_\theta\Delta v_\phi +
 {\cal O}\left( (\Delta v)^3 \right),\cr}
$$
it follows from equations (8-64) and (8-65) of Binney \& Tremaine (1987) that
$$\DRR=
R {64\pi^2r^2G^2m_\star^2\ln\Lambda\over 3J\c^2}
  \left(3I_{1\over2}-I_{3\over2}+2I_0\right)+{\cal O}(R^2).
$$
Notice that $\DR={1\over2}\p\DRR/\p R$, the orbit-averaged version of which
holds generally whenever the scattering is done by an external perturbation
(Landau 1937; Binney \& Lacey~1988).

\section {Numerical solution for the distribution function}

For each galaxy model, we need to find a
smooth DF $f(\E)$ that provides an acceptable fit to the model's
number-density profile $\nu(r)$ and potential $\psi(r)$ through
$$\nu(r)=4\pi\int_0^{\psi(r)}f(\E)\sqrt{2[\psi(r)-\E]}\,\d\E.
$$
The number density is given on a grid $(\log r_i,\log \nu(r_i))$ with
$n_\nu\sim60$ points, equispaced in $\log r$.  We represent $f(\E)$ on
a grid $(\log r(\E_i),\log f(\E_i))$ with $r(\E)$ defined such that
$\E=\psi(r(\E))$.  This grid has $n_f=1.5n_\nu$ points, with $r$
running logarithmically from the radius of the innermost $\nu(r)$
point to a few times the radius of the outermost one.  Values of
$f(\E)$ at intermediate points are obtained by linear interpolation in
$(\log r(\E),\log f(\E))$.

For a given trial $f(\E)$, it is straightforward to
integrate equation~\refeq1 numerically to obtain that model's density
distribution $\bar\nu(r)$.  We measure the goodness of fit of
each model using 
$$\chi^2\equiv\sum_{i=1}^{n_\nu}
\left(\nu_i-\bar{\nu}_i\over\nu_i\right)^2.
$$
Not all trial DFs that produce a low $\chi^2$ are equally acceptable
though, since some will be more jagged than others.  We penalize
jagged solutions using a penalty function that measures
the mean-square change in $\d f(\E)/\d r(\E)$:
$$P[f]\equiv {1\over n_f}\sum_{i=2}^{n_f-1}
\left({\log f(\E_{i+1})-2\log f(\E_i)+\log f(\E_{i-1})
\over \Delta\log r(\E)}\right)^2
$$
Then our most acceptable model is the one that maximizes the penalized
likelihood $\cal L$ defined through 
$$\log {\cal L}\equiv -{1\over2}\chi^2-\lambda P[f].
$$
We somewhat arbitrarily choose $\lambda$ such that a change in $P[f]$
of 10 causes the same change in $\cal L$ as a change in $\chi^2$ of
$10^{-4}n_\nu$ (a change in the RMS fractional error of 0.01).
Choosing different values for $\lambda$ causes insignificant changes
in the resulting DF.

Our procedure for maximizing equation~\refeq1 is as follows.  First we
fit an initial parameterized model of the form
$$f(\E)=A\left(r(\E)\over r_0\right)^\alpha\left(1+{r(\E)\over
r_0}\right)^\beta
$$
to the given $\nu(r)$.  We find the parameters $(A,r_0,\alpha,\beta)$
that minimize $\chi^2$ using the downhill-simplex method (Press et
al.\ 1992).  Then we use the Metropolis algorithm in the form
described by Magorrian (1999) to maximize~$\cal L$.  After a few
thousand iterations of the Metropolis algorithm, the model is still
acceptably smooth, with a typical RMS fractional error in $\nu(r)$ of
$0.01$ or better.

\bye